\documentclass[12pt]{article}

\usepackage{bbm}
\usepackage{bbold}
\usepackage{color}
\usepackage{authblk}
\usepackage{latexsym}
\usepackage{epsfig,amssymb,euscript, mathrsfs,cite}
\usepackage{amsmath}
\usepackage{slashed}
\usepackage{mathrsfs} 
\usepackage{enumerate}
\definecolor{MyDarkBlue}{rgb}{0.15,0.15,0.45}
\usepackage[linktocpage=true]{hyperref}
\hypersetup{
colorlinks=true,
citecolor=MyDarkBlue,
linkcolor=MyDarkBlue,
urlcolor=MyDarkBlue,
pdfauthor={},
pdftitle={},
pdfsubject={hep-th}
}
\usepackage{chngcntr}
\counterwithin*{equation}{section}

\newcommand{\ex}{\mathrm{e}}
\newcommand{\ii}{\mathrm{i}}
\newcommand{\ext}{\widetilde{\mathrm{e}}}
\newcommand\diff{\mathrm{d}}
\newcommand\Diff{\mathrm{D}}
\newcommand{\de}{\partial}
\newcommand{\vol}{\mathrm{vol}}

\newcommand{\R}{\mathbb{R}}

\newcommand{\Z}{\mathbb{Z}}
\newcommand{\N}{\mathbb{N}}
\newcommand{\cN}{\mathcal{N}}
\newcommand{\ds}{\diff s^2}

\newcommand{\pp}{p}
\newcommand{\qq}{q}
\newcommand{\rr}{s}

\newcommand{\spindle}{\mathbb{\Sigma}}
\newcommand{\disk}{\mathbb{D}}
\newcommand{\genus}{\mathtt{g}}
\newcommand{\sigmag}{\Sigma_{\mathtt{g}}}
\newcommand{\charge}{Q}
\newcommand{\ads}{\mathrm{AdS}}
\newcommand{\pflux}{\mathfrak{p}}
\newcommand{\nn}{\mathfrak{n}}
\newcommand{\hol}{\mathfrak{h}}

\newcommand{\UU}{V}
\newcommand{\VV}{U}

\newcommand{\gammatwo}{\gamma^{(\mathrm{2d})}}

\newcommand{\gammafour}{\gamma^{(\mathrm{4d})}}

\newcommand{\gammatwoE}{\tau^{(\mathrm{2d})}}

\newcommand{\gammathreeE}{\tau^{(\mathrm{3d})}}

\newcommand{\gammafourL}{\gamma^{(\mathrm{4d})}}
\newcommand{\gammafiveE}{\tau^{(\mathrm{5d})}}

\newcommand{\gammasixL}{\gamma^{(\mathrm{6d})}}

\newcommand{\gammasevenL}{\gamma^{(\mathrm{7d})}}
\newcommand{\gammaeightE}{\tau^{(\mathrm{8d})}}
\newcommand{\gammaeightL}{\gamma^{(\mathrm{8d})}}

\newcommand{\p}{p}
\newcommand{\q}{q}
\newcommand{\s}{s}

\newcommand{\es}[2] {\begin{equation} \label{#1} \begin{split} #2 \end{split} \end{equation}}

\begin{document}

\begin{titlepage}

\vskip 1cm

\begin{center}


{
\Large \bf A story of non-conformal branes: \\
\vspace{0.2cm}
spindles, disks, circles and black holes
}

\vskip 1cm
{Mathieu Boisvert$^{\mathrm{a}}$,  
and Pietro Ferrero$^{\mathrm{b}}$}

\vskip 0.5cm

${}^{\,\mathrm{a}}$\textit{C. N. Yang Institute for Theoretical Physics, \\
SUNY, Stony Brook, NY 11794, USA\\}

\vskip 0.2 cm

${}^{\mathrm{b}}$\textit{Simons Center for Geometry and Physics,\\
SUNY, Stony Brook, NY 11794, USA\\}

\end{center}

\vskip 0.5 cm

\begin{abstract}
\noindent  

We consider the $(p+2)$-dimensional gauged supergravities arising as a consistent truncation of type II on $S^{8-p}$, which are associated with the near-horizon limit of D$p$-branes, for $p=2,4,5,6$ (and NS5-branes for $p=5$). In a truncation of these theories with only abelian gauge fields and scalars, we find several classes of new solutions, with and without supersymmetry. Our ansatz for such backgrounds is inspired by the recent progress in the study of branes wrapped on orbifolds, but unlike those examples we consider ``non-conformal branes'', {\it i.e.} no Anti de Sitter factors in the metric. Focusing on cases with non-trivial gauge fields, we can divide the solutions that we present in three categories: 1) branes wrapping Riemann surfaces, spindles and disks, 2) branes wrapped on a circle with a holonomy for the gauge field along the circle and 3) electrically charged black holes in gauged supergravity, which uplift to rotating branes in ten dimensions. We carefully analyze the conditions for supersymmetry in all these cases.
\end{abstract}

\end{titlepage}

\pagestyle{plain}
\setcounter{page}{1}
\newcounter{bean}
\baselineskip18pt

\addtocontents{toc}{\protect\setcounter{tocdepth}{2}}

\tableofcontents

\newpage

\section{Introduction}\label{sec:intro}

Ever since the AdS/CFT correspondence was introduced in \cite{Maldacena:1997re} for extended M2, D3 and M5-branes in a flat transverse space, several variations of such simple setup have been considered. A particularly interesting one involves wrapping branes on compact cycles in special holonomy manifolds \cite{Maldacena:2000mw}, which provides the holographic description of the compactification of certain Quantum Field Theories (QFTs) on such cycles. Most of these constructions involve superconformal field theories (SCFTs) and the standard way to preserve supersymmetry in the compactification has been for a long time that of a partial topological twist \cite{Witten:1988ze}. In the case of two-dimensional cycles, such constructions typically involved genus-$\genus$ Riemann surfaces $\sigmag$ equipped with constant curvature metrics.

A new ingredient in this story was introduced in \cite{Ferrero:2020laf}, where it was shown that D3-branes can be wrapped on certain two-dimensional orbifolds called {\it spindles} $\spindle$, which is topologically a two-sphere with quantized conical deficits at the poles. Spindles do not admit metrics of constant curvature and moreover the way supersymmetry is preserved in this construction is not the standard topological twist. Rather, as later observed in \cite{Ferrero:2021etw}, there are two possibilities that were termed {\it twist} and {\it anti-twist}. Moreover, the conical deficits of the spindle are smoothed out almost everywhere by certain fibrations in the uplifted ten/eleven-dimensional supergravity background and the holographic matching of a central charge computation between field theory and gravity has been interpreted as a strong indication that branes wrapped on spindles provide {\it bona fide} examples of the AdS/CFT correspondence.

Following \cite{Ferrero:2020laf}, many examples of solutions describing branes wrapping orbifolds have appeared in the literature. Typically, these are found using the strategy first introduced in \cite{Maldacena:2000mw}: a solution is found in a certain gauged supergravity which is a consistent truncation of type IIA/IIB or 11d supergravity, and then it is uplifted on a suitable internal space. So far as spindles are concerned, solutions have been found describing D3 \cite{Ferrero:2020laf,Hosseini:2021fge,Boido:2021szx,Ferrero:2021etw}, M2 \cite{Ferrero:2020twa,Cassani:2021dwa,Ferrero:2021ovq,Ferrero:2021etw,Couzens:2021rlk,Couzens:2021cpk} and M5-branes \cite{Ferrero:2021wvk,Ferrero:2021etw} wrapping spindles, as well as D4-branes in the presence of D8 flavor branes \cite{Giri:2021xta,Faedo:2021nub} and D2-branes in the presence of various sets of other branes \cite{Couzens:2022yiv}. Various extensions have also been considered, including branes wrapped on the tensor product between a spindle and a 2d or 3d hyperbolic space of constant curvature \cite{Boido:2021szx,Faedo:2021nub,Giri:2021xta,Couzens:2022yiv,Suh:2022olh,Couzens:2022lvg,Couzens:2022aki,Cheung:2022ilc}, branes wrapped on four-dimensional orbifolds \cite{Faedo:2022rqx,Cheung:2022ilc,Bomans:2023ouw,Couzens:2022lvg,Faedo:2024upq,Macpherson:2024frt} and finally supergravity theories with hypermultiplets \cite{Arav:2022lzo,Suh:2022pkg,Suh:2023xse,Hristov:2023rel,Amariti:2023gcx,Amariti:2023mpg}. 

In \cite{Bah:2021mzw} it was also understood that in the same class of local solutions that describe spindles one can make a choice of parameters which doubles the supersymmetry and a different choice of global analysis leads to branes wrapped on disks $\disk$. These disks are 2d manifolds with an orbifold singularity at the origin and a curvature singularity at the boundary, which have been interpreted in terms of regular and irregular punctures, respectively, in the class-S description \cite{Gaiotto:2009we,Gaiotto:2009gz} of Argyres-Douglas theories \cite{Argyres:1995jj}. The disk construction resembles the spindle one in that the local supergravity solutions are literally the same, as first noticed in \cite{Couzens:2021tnv}, and the supersymmetry-preserving mechanism is again different from the topological twist. Solutions describing branes wrapping disks have also been considered for D3 \cite{Couzens:2021tnv,Suh:2021ifj,Suh:2021hef}, M2 \cite{Couzens:2021rlk}, D4-D8 \cite{Suh:2021aik} and M5-branes \cite{Bah:2021mzw,Bah:2021hei,Suh:2021ifj,Couzens:2022yjl,Karndumri:2022wpu}. See also \cite{Couzens:2023kyf} for a generalization that realizes the dual of class-S theories with orthogonal and symplectic gauge group.

A common feature of all the spindle and disk solutions described above is that they contain an AdS factor in the metric, which makes them dual to SCFTs. The presence of conformal symmetry is helpful in identifying the dual theory and performing explicit computations, but it has been known for a long time that the notion of holography extends beyond the realm of AdS/CFT and more generally relates gravitational theories to QFTs defined in one lower dimension \cite{Susskind:1994vu}. An example that will be relevant here is what was called domain wall/QFT correspondence in \cite{Boonstra:1998mp}. It has been shown that supergravity solutions describing branes wrapped on compact cycles can be found even in the absence of conformal invariance, still relying on a partial topological twist as a mechanism to preserve supersymmetry. The earliest constructions involve NS5/D5-branes wrapped on $S^2$ \cite{Maldacena:2000yy,Gauntlett:2001ps,Bigazzi:2001aj} in the attempt to find the holographic description of 4d $\mathcal{N}=1$ and $\mathcal{N}=2$ pure super Yang-Mills (SYM). Focusing on the case of two-cycles, generalizations to other types of D-branes wrapped on $S^2$ have been later found for D4-branes \cite{DiVecchia:2001uc}, other fivebrane solutions \cite{Casero:2006pt} and D6-branes \cite{Edelstein:2001pu} -- see also \cite{Paredes:2004xw} for considerations on the case of D5 and D6 branes wrapped on hyperbolic Riemann surfaces.

In this paper we consider a particularly simple ansatz which leads to new supersymmetric solutions describing D$p$-branes wrapped on Riemann surfaces, spindles and disks for $p=2,4,5,6$, inspired the analogous solutions for conformal branes. Moreover, we find that the new solutions introduced here admit a simple generalization by means of a parameter that we call $\ell$: for $\ell=1$ we recover spindle/disk solutions, while $\ell=0$ leads to completely different physics. As we shall see, this case is related to the recent constructions of solutions with branes wrapped on a circle, with a holonomy for the gauge fields along the circle \cite{Anabalon:2021tua,Anabalon:2022aig,Nunez:2023nnl,Nunez:2023xgl,Fatemiabhari:2024aua}. This type of solutions is dual to the compactification of certain dual theories on the circle, along which a Wilson line is extended.Finally, a double analytical continuation in the spirit of \cite{Lu:2003iv}, analogous to that relating the AdS black holes of \cite{Cvetic:1999xp} to spindle solutions, allows to use the same local metrics to access yet another type of physics: electrically charged black holes in gauged supergravity, which uplift to rotating $p$-branes in ten dimensions.

The rest of this paper is organized as follows. We begin with a general overview of the ideas behind our work, as well as a summary of our results, in Section \ref{sec:overview}. We then proceed considering a different type of brane (and the associated gauged supergravity) in each section: D6-branes in Section \ref{sec:D6}, D5- and NS5-branes in Section \ref{sec:D5}, D4-branes in Section \ref{sec:D4} and finally D2-branes in Section \ref{sec:D2}. For all these cases, we give the action and Killing spinor equations for the abelian truncation of gauged supergravity that we work with and present various new solutions (supersymmetric and non) describing branes wrapped on compact two-cycles, branes wrapped on a circle and rotating branes. We conclude in Section \ref{sec:discussion} with a discussion of our results and some outlook. Our work is complemented by three appendices, where we collect some technical comments and calculations: in Appendix \ref{app:gammas} we clarify our notation for gamma matrices, in Appendix \ref{app:7dsusy} we carefully investigate the structure of the Killing spinors preserved by our supersymmetric solutions in seven dimensions and finally in Appendix \ref{app:4dsugraderivation} we make some comments that are useful for the derivation of the action and Killing spinor equations of 4d $ISO(7)$ gauged supergravity.

\section{General overview}\label{sec:overview}

The core of this paper consists in presenting new solutions of gauged supergravities in various dimensions, all of which can be uplifted to certain configurations of D-branes in ten dimensions. The details of these backgrounds differ from case to case, but the underlying structure is similar. In this section, we give a general overview of the solutions that we present in the rest of the paper, explaining the ideas behind the ansatz that we make and emphasizing the common features.

\subsection{Gauged supergravities for non-conformal branes}\label{sec:supergravities}

Let us begin by listing here the gauged supergravities that we are interested in and highlighting the common features. As first explained in \cite{Boonstra:1998mp}, associated with each type of 1/2-BPS $p$-brane in type II supergravity\footnote{The same holds, of course, for membranes in eleven dimensions. However, since both M2 and M5-branes support conformal field theories on their worldvolume, those cases have already been thoroughly investigated in the literature and we shall not consider them here.} there is a consistent truncation on the $S^{8-p}$ transverse to the brane, to a $d=(p+2)$-dimensional gauged supergravity. The compact part of the gauge group is always $SO(9-p)$, gauging the isometries of $S^{8-p}$, but a non-compact part is also possible. In the case of branes which support conformal field theories, the near-horizon limit is of the form $\ads\times S$ and there is a supersymmetry enhancement in the limit, which implies the existence of a maximally supersymmetric AdS vacuum in the corresponding gauged supergravity. This is the situation for D3-branes, as well as for M2- and M5-branes in eleven dimensions. On the contrary, there is no supersymmetry enhancement in the near-horizon limit of ``non-conformal'' branes, which implies the absence of maximally supersymmetric vacua in the corresponding gauged supergravities. Rather, in all these theories there is a special 1/2-BPS supersymmetric vacuum which is a domain wall supported by a single scalar preserving the full $SO(9-p)$ isometry of the internal sphere, as well as a family of 1/2-BPS vacua which break (partially or entirely) $SO(9-p)$. The former corresponds to the near-horizon region of coincident branes, while the latter uplifts to the decoupling limit of certain continuous distributions of branes \cite{Sfetsos:1998xd,Kraus:1998hv,Freedman:1999gk,Bakas:1999ax,Cvetic:1999xx,Bakas:1999fa,Cvetic:2000zu,Bakas:2000nt}.

The consistent truncation of type II on $S^{8-p}$ gives rise to several bosonic fields in gauged supergravity: the metric, of course, as well as differential forms of various degrees, which can be classified as Neveu–Schwarz-Neveu–Schwarz (NSNS) or Ramond-Ramond (RR) according to their higher-dimensional origin. In the following we are going to be interested in a truncation of such theories where all RR fields of the $d=(p+2)$-dimensional gauged supergravity are set to zero, while among the NSNS fields we focus on i) $n=\lfloor \tfrac{9-p}{2}\rfloor$ massless vectors which gauge the maximal abelian subgroup $U(1)^n\subset SO(9-p)$ and ii) the scalars that are singlets of $U(1)^n$, and are therefore not charged under the gauge fields that we consider\footnote{The case $p=6$ is special in this sense: there is only one gauge field, which however sources the RR one-form potential in type IIA and not the metric, as opposed to all other cases.}. Let us summarize here the theories associated with each type of branes that we consider, which we stress again are all maximally supersymmetric.
\begin{itemize}
\item {\bf D6-branes:} the reduction of type IIA on $S^2$ gives rise to an 8d $SU(2)$-gauged theory \cite{Salam:1984ft}. We are interested in its truncation to a sector with an abelian gauge field for the Cartan of $SU(2)$ and two scalars, which can be thought of as associated with the dilaton and a volume modulus for the $S^2$, respectively. This is the truncation first considered in \cite{Edelstein:2001pu}.

\item {\bf D5-branes:} the reduction of type IIB on $S^3$ gives a 7d $SO(4)$-gauged supergravity, as first discussed in \cite{Samtleben:2005bp}. We are interested in a truncation with two gauge fields for the Cartans of $SO(4)$, as well as two scalar fields. This truncation was first considered in \cite{Gauntlett:2001ps,Bigazzi:2001aj}.

\item {\bf NS5-branes:} here we should distinguish between the IIA and IIB NS5-branes. In the former case, the reduction on $S^3$ gives a 7d $ISO(4)$ gauged supergravity, which can be thought of as a limit of the theory with maximal $SO(5)$ gauging \cite{Cvetic:2000ah}. On the other hand, the type IIB NS5-brane is S-dual to the D5-brane, so the truncation is the same $SO(4)$-gauged theory discussed in that case. Moreover, the $ISO(4)$ and the $SO(4)$ theory only differ in the RR sector, so for our purposes we will consider the same truncation for all three cases: D5-, IIA NS5- and IIB NS5-branes.

\item {\bf D4-branes:} the consistent truncation of type IIA on $S^4$ is a 6d $SO(5)$ gauged theory which, as first discussed in \cite{Cowdall:1998rs}, arises from the circle reduction of the more famous 7d maximal $SO(5)$ gauged supergravity \cite{Pernici:1984xx}, {\it i.e.} the consistent truncation associated with M5-branes. We are interested in the truncation to a sector with two abelian gauge fields and three scalars, which arises from the circle reduction of the truncation of the 7d $SO(5)$ theory first considered in \cite{Liu:1999ai}.

\item {\bf D2-branes:} the reduction of IIA on $S^6$ gives a 4d theory with non-compact $ISO(7)$ gauging, which can be thought of as a limit of the theory with maximal $SO(8)$ gauging \cite{Hull:1984yy}. We consider the truncation to three abelian gauge fields (the Cartans of $U(1)^3\subset SO(7)\subset ISO(7)$) and four scalars. This is the analogue of the truncation first considered in \cite{Duff:1999gh} for the maximal $SO(8)$ gauging, with the difference that it has one fewer gauge field (since one of the Cartans of $SO(8)$ becomes non-compact in the $ISO(7)$ limit) and one additional scalar (the dilaton after uplift to type IIA). Since this specific truncation has not previously appeared in the literature we shall give some details on its derivation in Appendix \ref{app:4dsugraderivation}.
\end{itemize}

In all cases, we shall focus on solutions with non-trivial gauge fields, which after the uplift are connections for a fibration of the internal $S^{8-p}$ on the $(p+2)$-dimensional spacetime of the gauged supergravity.\footnote{Again, with the exception of D6 branes, where the 8d gauge field sources the RR one-form potential in type IIA.} Focusing on cases where a certain amount of supersymmetry is preserved and with independent non-trivial charges for all gauge fields, the number of Killing spinors doubles every time one of the gauge fields is set to zero, since that trivializes one part of the fibration.

\subsection{The idea behind the ansatz}\label{sec:philosophy}

The original motivation for our work is to find solutions describing ``non-conformal branes'' wrapped on two-cycles $\Sigma$, generalizing what is known for the ``conformal'' cases which have been subject of thorough investigations in the literature, starting with \cite{Maldacena:2000mw}. When the dual theory is conformal, one can imagine a putative ``full-flow'' solution in $d$-dimensional gauged supergravity\footnote{Such solutions are typically only known numerically. The simplest case, where analytic expressions are available, is that of extremal black holes in $\ads_4$, where the coordinate that we are referring to as $r$ here is the radial coordinate of the black hole.}, with a radial coordinate $r$ interpolating between $\ads_{d+1}$ at $r=+\infty$ and $\ads_{d-1}\times \Sigma$ at, say, $r=0$. Since full-flow solutions are typically hard to find, many authors have focused on the study of the near-horizon regime, where the dependence on $r$ is completely frozen by the isometries of AdS. When $\Sigma$ is a genus-$\genus$ Riemann surfaces $\sigmag$ \cite{Cacciatori:2009iz,Bah:2012dg,Benini:2013cda,Karndumri:2015eta}, the metric is completely fixed by isometries up to some constants. More recently, starting with \cite{Ferrero:2020laf} near-horizon solutions with branes wrapped on orbifolds called spindles $\spindle$ have been explored, see \cite{Ferrero:2021etw} for a general overview. Spindles are equipped with co-homogeneity one metrics depending on two coordinates traditionally called $y$ and $z$: $z$ parametrizes a $U(1)$ direction, while the metric depends non-trivially on $y$. The metric on spindles is then usually written as 
\es{generalspindlemetric}{
\ds=\frac{y^{\#}\diff y^2}{P}+\frac{P}{H}\diff z^2\,,
}
where, according to the construction where the spindle is embedded, for a certain value of $\#$ the functions $P$ and $H$ are polynomials in $y$. The problem of finding near-horizon $\ads_{d-1}\times \spindle$ solutions involves solving ordinary differential equations (ODEs) in $y$. 

Here we are interested in the case of non-conformal branes, that is solutions {\it without} AdS factors. The dependence on $r$ is no longer fixed by isometries, so {\it a priori} the problem is significantly more complicated: naively, one should solve ODEs in $r$ to find solutions describing branes wrapped on Riemann surfaces, and partial differential equations (PDEs) in $(r,y)$ for branes wrapped on spindles. Examples of these solutions are so far available only for the case of Riemann surfaces: D6-branes wrapped on $S^2$ which uplift to the conifold in 11d \cite{Edelstein:2001pu}, D5-branes wrapped on $S^2$ with non-abelian gauge fields\footnote{Solutions in the abelian version of the theory were also found but considered too singular to be relevant: the singularity is resolved by considering non-abelian gauge fields.} for the dual of 4d $\cN=1$ SYM \cite{Maldacena:2000yy} and abelian solutions with doubled supersymmetry for the dual of 4d $\cN=2$ SYM \cite{Gauntlett:2001ps,Bigazzi:2001aj}. In all these solutions, the dependence of the various functions on the $S^2$ coordinates is fixed by isometries and all functions depend non-trivially on one coordinate ($r$, in the previous paragraph). 

To find new solutions involving Riemann surfaces, as well as spindle solutions which are so far unavailable, we are going to rely on the observation of \cite{Boonstra:1998mp} that the near-horizon region of D$p$-branes is actually $\ads_{p+2} \times S^{8-p}$ up to a Weyl rescaling by the dilaton: there is in fact a symmetry at play, involving a transformation of the dilaton (which is not constant) as well as the metric, which we can use to ``freeze'' the $r$-dependence, thus making the problem analogous to that of AdS solutions in terms of differential equations.\footnote{As noted in \cite{Boonstra:1998mp} the case of fivebranes is special, but a similar mechanism can be invoked.} It is of course possible to imagine that ``full-flow'' solutions should exist in this case too, like in the AdS analogues, which cure some of the pathologies of the simpler ones that we consider here. A famous example of this is the singularity of the linear-dilaton solution representing the near-horizon limit of coincident NS5-branes, which can be cured by separating the branes and taking a decoupling limit, after which one obtains the so-called ``cigar CFT'' \cite{PhysRevD.44.314,Sfetsos:1998xd}: here the dependence on the equivalent of $r$ is more complicated, but the solution is less singular. However, we shall not consider such full-flow solutions here and focus instead on finding new near-horizon solutions for the branes listed in the previous subsection, filling a gap in the literature.

\subsection{Summary of results}\label{sec:summary}

Since our initial motivation was to obtain spindle solutions, that will always be our starting point. We shall not review spindles here and we refer the reader to the references cited in the Introduction for an overview of the literature. Once spindle solutions are obtained, however, several minor modifications are possible leading to other interesting solutions, which we also describe in this paper. Let us summarize here all the types of solutions that we explore as well as presenting our results in a schematic way.

Leaving aside the case of fivebranes, which is somewhat special, the generic form that we find for solutions describing $p$-branes wrapped on a two-cycle $\Sigma$ is, schematically\footnote{In the case of fivebranes one has to consider a different ansatz, as it will become clear in Section \ref{sec:D5}, but a completely analogous story holds in that case too so we do not comment on that separately here.}
\es{ansatz_generic_ell1}{
\ds_{p+2}\sim r^{\#}\,\left[\ds(\R^{1,p-2})+\diff r^2+r^2\,\ds(\Sigma)\right]\,,
}
where we have emphasized the dependence on $r$ and $\#$ is a certain rational number fixed by $p$. This ansatz works when $\Sigma\equiv \sigmag$ is a Riemann surface of genus $\genus$, a spindle $\spindle$ or a disk $\disk$ of the type introduced in \cite{Bah:2021mzw} (see the Introduction for references). Of course, the identification of \eqref{generalspindlemetric} with a metric on a spindle or on a disk relies on a certain choice of compact range for the coordinate $y$: we find that this is always possible, but other choices might be also physically interesting. In the case of AdS solutions for instance, the same local metrics that give spindles can also be used to study conformal defects and Rényi entropies holographically \cite{Huang:2014gca,Nishioka:2014mwa,Bianchi:2016xvf,Hosseini:2019and} by choosing a non-compact range for $y$. We do not look into that possibility here, but it would be interesting to do so in the future. 

We would also like to emhpasize that this type of ansatz generically gives rise to singular backgrounds at $r=0$ (or $\rho=-\infty$ in the case of fivebranes, see Section \ref{sec:D5}). This does not automatically make our solutions uninteresting: rather, they should considered as a stepping stone towards more general and regular solutions. As to how such more regular solutions would look like, we wish to mention to notable examples. The first is the near-horizon solution of NS5 branes: the dilaton is linear in one direction ($\rho$), with a negative coefficient, so that $\rho\to -\infty$ gives a divergent string coupling. This, famously, can be cured replacing the linear dilaton with a geometry with more complicated dependence on $\rho$, commonly referred to as the Euclidean black hole, or the cigar \cite{PhysRevD.44.314}, which asymptotes the linear dilaton in the region of small string coupling but fixes its singularity with smoothly capped off geometry at $\rho=0$. We take this as a hint of the fact that an ansatz with a general dependence on both $r$ and the coordinates on $\Sigma$ in \eqref{ansatz_generic_ell1} is likely to lead to more regular solutions, but this would imply (in general) dealing with partial differential equations, thus leading to a significantly more complicated problem. In the case where $\Sigma$ is a Riemann surface the problem is more tractable, and indeed it was solved in the cases of D5 and D6 branes wrapped on $S^2$ in \cite{Gauntlett:2001ps,Bigazzi:2001aj} and \cite{Edelstein:2001pu}, respectively. We do not try and solve the analogous problems for general D$p$ branes wrapped on spindles here, but we refer the reader to \cite{Ferrero:2024vmz} for comments on the case of D6 branes, where this is related to certain known metrics for the small resolution of the cone over $Y^{p,q}$ manifolds \cite{Martelli:2007pv,Martelli:2007mk}, showing that indeed more general solutions depending non-trivially on two variables exist and turn out to be less singular. It will be therefore worth studying the analogous problem for general D$p$ branes in the future.

Another interesting approach that could lead to more general solutions involves, on top of allowing for a more general dependence on $r$, the inclusion of non-abelian gauge fields. Indeed, in this paper we always focus on abelian truncations of gauged supergravities where a non-abelian group is gauged, and it is quite possible that interesting solutions would involve a more general choice of vectors. This was for instance the case, famously, for the dual of $\mathcal{N}=1$ SYM in \cite{Maldacena:2000mw}, where the abelian truncation of gauged supergravity leads to a solution which was quickly discarded by the authors, in favor of a more complicated -- yet more interesting -- solutions involving non-trivial $SU(2)$ gauge fields twisting a two-sphere. This is certainly a promising, but even more complicated direction, that we do not consider here. 

An interesting observation that we make in this paper is that a simple modification of the ansatz \eqref{ansatz_generic_ell1}, namely
\es{ansatz_generic_ell}{
\ds_{p+2}\sim r^{\#\,\ell}\,\left[\ds(\R^{1,p-2})+r^{2(\ell-1)}\diff r^2+r^{2\ell}\,\ds(\Sigma)\right]\,,
}
which reduces to \eqref{ansatz_generic_ell1} for $\ell=1$, also gives interesting solutions when $\ell=0$. In that case, the dependence on $r$ drops from all the fields and $\log r$ becomes a flat direction, so that one can effectively replace $\ds(\R^{1,p-2})+r^{-2}\diff r^2\to \ds(\R^{1,p-1})$. On the other hand, we find that in all cases it is not possible to choose a compact range for $y$ such that the metric is real, has the correct signature and the $z$-circle shrinks at the extrema of the interval. Thus, $\ds(\Sigma)$ should not be thought of as a compact two-cycle: rather, the solution should be interpreted in the spirit of \cite{Anabalon:2021tua,Anabalon:2022aig,Nunez:2023nnl,Nunez:2023xgl,Fatemiabhari:2024aua}, as the dual of a QFT compactified on a circle with a Wilson line along the circle. In this case one considers the largest root of $P$, say $y=y_+$, and a non-compact range $y\in(y_+,+\infty)$ is chosen. The periodicity of $z$ can be chosen in such a way that the circle it parametrizes shrinks smoothly for $y\simeq y_+$, while as $y\to +\infty$ we recover the ``vacuum'' of these theories: a domain wall which uplifts to the near-horizon region of a stack of branes in flat space, with one of the worldvolume directions compactified and $y$ playing the role of an energy scale.

Finally, it is always possible to perform a double analytical continuation in the spirit of \cite{Lu:2003iv}, exchanging the role of the time direction with that of $z$ in \eqref{generalspindlemetric}. For $\ell=0$, this leads to electrically charged black holes in gauged supergravity with a flat horizon, which uplift to rotating branes in ten dimensions (the rotation parameters being provided by the electric charges): solutions of this type were studied from a ten-dimensional viewpoint in \cite{Cvetic:1996dt,Cvetic:1999xp,Harmark:1999xt,Sfetsos:1999pq,Nakayama:2008kg}, for example.

In the rest of the paper we go through the list of supergravities of Section \ref{sec:supergravities} and find, in each case, solutions of the type described above: 
\begin{itemize}
\item[i)] branes wrapped on compact two-cycles, {\it i.e.} solutions coming from the ansatz \eqref{ansatz_generic_ell} with $\ell=1$. Here $\Sigma$ could be either a Riemann surface $\sigmag$, a spindle $\spindle$ or a disk $\disk$, and we will show which choice of coordinates and parameters realizes each case. 
\item[ii)] branes wrapped on a circle, {\it i.e.} solutions coming from the ansatz \eqref{ansatz_generic_ell} with $\ell=0$. In this case the circle is the only compact direction.
\item[iii)] rotating branes, {\it i.e.} solutions coming from a double analytical continuation of the ansatz \eqref{ansatz_generic_ell} with $\ell=0$, which are electrically charged black holes in gauged supergravity.
\end{itemize}
In each dimension we will start from a ``general solution'', depending on $\ell$, and then fixing $\ell=0,1$ or performing the analytical continuation we shall make comments about specific solutions, in particular stating which solutions preserve supersymmetry and, if they do, which fraction of the supersymmetry is preserved. This comes from an explicit derivation of the corresponding Killing spinors, but for brevity we only present the result of this computation for the case of fivebranes, in Appendix \ref{app:7dsusy}.

Let us summarize our findings here, for the reader's convenience. We choose to highlight three main features: a) in the case of solutions describing branes wrapped on a Riemann surface $\sigmag$, for which curvature solutions exist, b) for spindle solutions (which are always supersymmetric), which twist is realized \cite{Ferrero:2021etw} and c) whether supersymmetric solutions exist in the other cases (either for branes wrapped on a circle or for black holes). Our results are as follows:
\begin{itemize}
\item[a)] {\bf Branes wrapped on Riemann surfaces.} We find that an ansatz of the type \eqref{ansatz_generic_ell1}, where $\Sigma\equiv \sigmag$ is a Riemann surface of genus $\genus$ equipped with a metric of constant curvature, always gives supersymmetric solution for at least one value of the curvature $\kappa$ ($\kappa=+1$ for $\genus=0$, $\kappa=0$ for $\genus=1$ and $\kappa=-1$ for $\genus>1$). The results for D$p$-branes wrapped on $\sigmag$ are summarized in Table \ref{tab:riemann_summary}, where ``yes'' means that solutions of the bosonic equations of motion exist, while ``no'' means that they do not. Moreover, as we emphasize in the table, whenever a solution exists it also always admits a supersymmetric limit. 
\begin{table}[h!]
\centering
\begin{tabular}{|c||c|c|c|c|} 
 \hline
  & $p=2$ & $p=4$ & $p=5$ & $p=6$\\
  \hline \hline
 $\kappa=+1$ & yes, susy & yes, susy & yes, susy & yes, susy \\
 \hline
 $\kappa=0$ & yes, susy & yes, susy & no & no \\
 \hline
 $\kappa=-1$ & yes, susy & yes, susy & no & no \\
 \hline
\end{tabular}
\caption{Summary of results for solutions describing D$p$-branes wrapped on a Riemann surface $\sigmag$, in terms of the curvature $\kappa$. }
\label{tab:riemann_summary}
\end{table}

\item[b)] {\bf Branes wrapped on spindles.} We find that solutions describing $p$-branes wrapped on spindles are possible for all values of $p$ and moreover they always admit a supersymmetric limit where all gauge fields in the abelian truncations of gauged supergravity that we consider have independent charge parameters. We remind that a spindle $\spindle$ is a weighted projective space $\mathbb{WCP}^1_{\nn_1,\nn_2}$, characterized by two co-prime integers $\nn_{1,2}$. Topologically, this is a two-sphere with quantized conical deficits $2\pi(1-\nn_i^{-1})$ at the two poles. An interesting quantity that characterizes spindle solutions is the flux $Q^{(R)}$ of the graviphoton gauge field across $\spindle$, which for supersymmetric solutions can only take two values \cite{Ferrero:2021etw}:
\es{}{
Q^{(R)}=
\begin{cases}
\frac{\nn_1+\nn_2}{\nn_1\,\nn_2}\,,\quad \text{twist}\,,\\
\frac{\nn_1-\nn_2}{\nn_1\,\nn_2}\,,\quad \text{anti-twist}\,.
\end{cases}
}
Some of the solutions that have appeared in the literature realize both types of twist, according to certain choices for the values of the parameters, while other solutions only realize one of the two. It is therefore interesting to summarize the twist realized by the spindle solutions that we investigate in this paper and we do so in Table \ref{tab:twists}.
\begin{table}[h!]
\centering
\begin{tabular}{|c|c|c|c|} 
 \hline
 $p=2$ & $p=4$ & $p=5$ & $p=6$\\
  \hline \hline
 twist & twist & anti-twist & twist\\
 \hline
\end{tabular}
\caption{Twist realize by gauged supergravity solutions describing D$p$-branes wrapped on a spindle. }
\label{tab:twists}
\end{table}
With our results, solutions describing a single type of either D$p$-branes, or NS5-branes, or membranes, wrapped on a spindle have now been found for all types of branes. As we do here, these are typically found in gauged supergravity and uplifted to either ten dimension (in the former two cases) or to eleven dimension (for membranes), where the internal space is a (squashed and fibered) sphere $S^n$. Let us leave the case of D6-branes aside for a moment. In all other cases, it seems that the existence of anti-twist solutions is associated with the possibility of taking a limit where all gauge fields are identified, whereas twist solutions require at least two distinct gauge fields. The only solutions where more than one type of twist is possible are those for M2- and D3-branes \cite{Ferrero:2021etw}, but the anti-twist is more ``universal'' in those cases, in that the existence of the twist requires multiple distinct gauge fields. With this caveat, one can argue from the study of the solutions found so far in the literature and here that the ``universal'' (in the sense described above) twist for spindle solutions seems to be the anti-twist for odd $n$ and the twist for even $n$ (the dimension of the internal $S^n$). Moreover, the uplift on odd-dimensional spheres smoothens the conical singularities of the spindle, giving rise to regular solutions in ten or eleven dimensions \cite{Ferrero:2020laf}, while conical singularities at the poles of $S^n$ persist when the uplift is performed on even-dimensional spheres, as first observed in \cite{Ferrero:2021wvk}. Going back to the case of D6-branes, we can see that it is somewhat special in that it has only one gauge field, which however in ten dimensions does not fiber the internal $S^2$ but rather sources the RR one-form. In that case the twist is realized, compatibly with the comment that the uplift on even-dimensional spheres should give twist solutions. On the other hand, it provides an example with only one gauge field which realizes the twist.

\item[c)] {\bf Other supersymmetric solutions.} While in all cases we are able to find solutions to the equations of motion that describe the other two physical setups that we consider here, namely branes wrapped on a circle and electric black holes/rotating branes, we find that almost all solutions of those types break supersymmetry completely. The only exception is that of fivebranes, where we find non-trivial supersymmetric solutions describing fivebranes wrapped on a circle, which generalize those of \cite{Nunez:2023xgl}. Moreover, we also find one supersymmetric solution which describes a black hole in seven dimensions, but it contains a naked singularity: more interesting black holes, which uplift to rotating fivebranes, are not supersymmetric.

\end{itemize}

\section{D6-branes}\label{sec:D6}

Supersymmetric solutions describing D6-branes wrapped on $S^2$ in the Eguchi-Hanson space $T^*S^2$ were found in \cite{Edelstein:2001pu}, both in the near-horizon ansatz \eqref{ansatz_generic_ell} (with $\ell=1$) and full-flow solutions with a more complicated dependence on $r$. Once uplifted to eleven dimensions, they give a background $\R^{1,4}\times \text{CY}_3$, where the Calabi-Yau three-fold $\text{CY}_3$ is the conifold in the case of the ansatz \eqref{ansatz_generic_ell} and its small resolution (obtained by blowing up a two-cycle) in the case of the full-flow solution. We consider this eleven-dimensional perspective for the case of spindle solutions in a separate paper \cite{Ferrero:2024vmz}, while here we focus on the D6-branes perspective.

\subsection{8d gauged supergravity}\label{sec:8dsugra}

We consider the 8d maximal $SU(2)$-gauged supergravity constructed in \cite{Salam:1984ft} and focus on the truncation discussed in \cite{Edelstein:2001pu}, which keeps only the metric, a gauge field $A$ (with $F=\diff A$) and two scalars $\phi$ and $\lambda$. The action for the bosonic fields is
\es{8daction}{
S_{\mathrm{8d}}=\frac{1}{16\pi G^{(8)}_N}\int \diff^8x\sqrt{-g}\left[R+\mathcal{V}-\frac{1}{2}(\partial\phi)^2-6(\partial\lambda)^2-\frac{1}{4}\ex^{\phi+4\lambda}(F_{\mu\nu})^2\right]\,,
}
where 
\es{V8d}{
\mathcal{V}=\frac{g^2}{2}\ex^{2\lambda-\phi}(4-\ex^{6\lambda})\,,
}
where $g$ is the gauge coupling. A solution to this model preserves supersymmetry if the following Killing spinor equations 
\es{KSE}{
\delta\psi_\mu=&\left[\nabla_\mu-	\frac{\ii}{2}g\, A_{\mu}\,\gammathreeE_3+\frac{1}{48}\ex^{2\lambda+\phi/2}\left(\gammaeightL_{\mu\nu\rho}-10 \delta_{\mu}^{\,\,\nu}\gammaeightL_\rho\right)\gammaeightL_\star\,F^{\nu\rho}\,\gammathreeE_3\right.\\
&\left.\,\,\,+\frac{\ii}{24}\,g\,\ex^{-\phi/2}(\ex^{4\lambda}+2\ex^{-2\lambda})\gammaeightL_\mu\,\gammaeightL_\star\right]\epsilon=0\,,\\
\delta\chi=&\left[\slashed{\partial}\lambda+\frac{1}{6}\slashed{\partial}\phi-\frac{\ii}{4}\,g\,\ex^{-\phi/2}(\ex^{4\lambda}-\ex^{-2\lambda})\,\gammaeightL_\star-\frac{1}{8}\ex^{2\lambda+\phi/2}\slashed{F}\,\gammaeightL_\star\,\gammathreeE_3\right]\epsilon=0\,,
}
are satisfied by non-trivial spinors $\epsilon$, where $\psi_{\mu}$ is the gravitino while $\chi$ is a matter fermion. Note that $\epsilon$ is a spinor of both the Lorentz group $Spin(1,7)$ and of the R-symmetry group $SU(2)\simeq SO(3)/\Z_2$, whose gamma matrices we denote with $\gammathreeE$.

The 8d supergravity of \cite{Salam:1984ft} was originally found as a Scherk–Schwarz compactification of 11d supergravity on $S^3$. As such, its (supersymmetric) solutions uplift to (supersymmetric) solutions of 11d supergravity. The specific truncation we are interested in is such that solutions to \eqref{8daction} uplift to pure geometry in 11d, with metric given by
\es{uplift8dto11d}{
\ds_{11}=\ex^{-\phi/3}\ds_8+\frac{1}{g^2}\ex^{2\phi/3}\left[\ex^{-2\lambda}(\diff\theta^2+\sin^2\theta\diff\alpha^2)+\ex^{4\lambda}(\diff\psi-\cos\theta\diff\alpha+g\,A)^2\right]\,,
}
where $(\theta,\alpha,\psi)$ parameterize $S^3$ if $\theta\in[0,\pi]$, $\alpha\in[0,2\pi)$ and $\psi\in[0,4\pi)$. Note that this can be reduced along the $\psi$-circle to a solution of type IIA supergravity with metric, dilaton and RR one-form $C_1$, which in the string frame read
\es{uplift8dto10d}{
\ds_{10}=&\ex^{2\lambda}\ds_8+\frac{1}{g^2}\ex^{\phi}(\diff\theta^2+\sin^2\theta\diff\alpha^2)\,,\\
C_1=&-\cos\theta\diff\alpha+g\,A\,,\\
\Phi=&\frac{\phi}{2}+2\lambda-\frac{1}{2}\log g\,,
}
which is the form of a IIA solution describing D6-branes. Indeed, the IIA solution describing the near-horizon limit of a stack of D6-branes can be found from the uplift of a solution to \eqref{8daction} whose only non-trivial fields are the metric and the scalar $\phi$, given by
\es{D6NH_8d}{
\ds_8&=r\,\left[\ds(\R^{1,6})+\diff r^2\right]\,,\\
\ex^{\phi}&=\frac{g^2\,r^3}{4}\,.
}

\subsection{The general solution}\label{sec:8dgeneral}

We start from the ansatz\footnote{Note that we have chosen a gauge such that the Killing spinors are uncharged under $\partial_z$.}
\es{ansatzwrappedD6}{
\ds_8&=r^{\ell}y^{1/6}H^{1/6}\left[\ds(\R^{1,4})+r^{2(\ell-1)}\diff r^2+r^{2\ell}\left(\frac{y\,\diff y^2}{g^2\,P}+\frac{g^2\,P}{4\,H}\diff z^2\right)\right]\,,\\
A&=\left(\frac{\s}{H}+\frac{\ell}{2}\right)\diff z\,,\\
\ex^{2\phi}&=\p_1^2\,r^{6\ell}\,y\,H\,,\quad
\ex^{6\lambda}=\frac{\p_2^3\,H}{y^2}\,,
}
where the function $H$ is
\es{H8dgeneral}{
H=y^2+\q\,,
}
and we have free parameters $\p_1$, $\p_2$, $\s$ and $\q$.The equations of motion fix
\es{P8dgeneral}{
P=\left(\frac{1}{\p_1}-4\frac{\ell^2}{g^2}y\right)H+\frac{4\p_1\s^2}{g^2\,\q}y\,,\quad \p_2=1\,,
}
and for both $\ell=0$ and $\ell=1$ we can invoke a scaling symmetry to fix
\es{}{
p_1=1\,.
}

\subsection{D6-branes wrapped on two-cycles}\label{sec:8dorbifolds}

\subsubsection{Warm up: Riemann surfaces}\label{sec:8driemann}

Let us begin with the simplest case of branes wrapped on a Riemann surface. We make the ansatz
\es{}{
\ds_8&=H_0^2\,r\,\left[\ds(\R^{1,4})+\diff r^2+r^2\,R^2\,\ds(\sigmag)\right]\,,\\
F&=\diff A=s\,\vol(\sigmag)\,,\quad
\ex^{2\phi}=p_1^2\,r^6\,,\quad
\ex^{6\lambda}=p_2^3\,,
}
in terms of the parameters $p_1$, $p_2$, $s$, $H_0$ and $R$. This ansatz can be obtained from a suitable scaling limit of \eqref{ansatzwrappedD6}, as explained in Appendix A of \cite{Ferrero:2021etw}. One can look for solutions of the equations of motion separately in the three cases $\genus=0$, $\genus=1$ and $\genus>1$, but it turns out that real solutions are possible only for $\genus=0$, that is D6-branes wrapped on $S^2$ -- see \cite{Paredes:2004xw} for comments on the hyperbolic case. In that case, a supersymmetric limit is also possible and we shall present the supersymmetric solution here, which is just the one found in \cite{Edelstein:2001pu}. We find the constraints
\es{}{
p_1=\frac{1}{6}\left(\frac{2}{3}\right)^{1/3}H_0^2\,g^2\,,\quad
p_2=\left(\frac{2}{3}\right)^{1/3}\,,\quad
R=\frac{1}{\sqrt{6}}\,,
}
where only the combination $H_0^2/p_1$ is physical so we can choose, for instance, $p_1=1$, while the charge parameter $s$ is fixed to be
\es{}{
s=\frac{1}{g}\,,
}
which is the topological twist condition.

\subsubsection{Spindles}\label{sec:8dspindles}

Let us now focus on the case $\ell=1$ in \eqref{ansatzwrappedD6} (with $p_1=1$) and consider supersymmetric solutions, which requires the condition
\es{}{
s=q\,.
}
The supersymmetric solution can then be written as
\es{8dsusyspindle}{
\ds_8&=r\,y^{1/6}H^{1/6}\left[\ds(\R^{1,4})+\diff r^2+r^{2}\left(\frac{y\,\diff y^2}{g^2\,P}+\frac{g^2\,P}{4\,H}\diff z^2\right)\right]\,,\\
A&=\left(\frac{q}{H}+\frac{1}{2}\right)\diff z\,,\\
\ex^{2\phi}&=r^{6}\,y\,H\,,\quad
\ex^{6\lambda}=\frac{H}{y^2}\,,
}
where the functions $H$ and $P$ are given by
\es{}{
H=y^2+q\,,\quad
P=H-\frac{4}{g^2}y^3\,,
}
in terms of the single parameter $q$. We study this solution in detail in \cite{Ferrero:2024vmz}, so we will only summarize some of our results here. Our goal is to prove that we can choose values for $q$, a compact range for $y$ and a periodicity for $z$ such that the space parametrized by $(y,z)$ is a spindle. As shown in \cite{Ferrero:2024vmz}, if we choose
\es{spindlepar8d}{
q&=-\frac{g^4}{216}\left(1-\frac{(\nn_1^2-4\nn_1\nn_2+\nn_2^2)}{(\nn_1+\nn_2)^3}\sqrt{\nn_1^2+14\nn_1\nn_2+\nn_2^2}\right)\,,\\
\Delta z&=\frac{2\pi}{9g\,\nn_1\nn_2(\nn_1-\nn_2)}\left[(\nn_1^2-10\nn_1\nn_2+\nn_2^2)-(\nn_1+\nn_2)\sqrt{\nn_1^2+14\nn_1\nn_2+\nn_2^2}\right]\,,
}
we obtain that all the conditions for global regularity, reality and Lorentzian signature of the metric are met if we choose $\nn_2>\nn_1$, where $\nn_{1,2}\in \N$ and hcf($\nn_1,\nn_2$)=1. In this case the polynomial $P$ has three real roots $y_0<y_1<y_2$ which are given in terms of the integers $\nn_{1,2}$ via
\es{spindleroots8d}{
y_1&=\frac{g^2}{4}\frac{\nn_1}{\nn_1+\nn_2}-\frac{1}{2}y_0\,,\quad
y_2=\frac{g^2}{4}\frac{\nn_2}{\nn_1+\nn_2}-\frac{1}{2}y_0\,,\\
y_0&=\frac{g^2}{12}\left(1-\frac{\sqrt{\nn_1^2+14\nn_1\nn_2+\nn_2^2}}{\nn_1+\nn_2}\right)\,.
}
It is straightforward to compute the total flux of the gauge field $A$ across the spindle, which reveals which twist is realized by this solution \cite{Ferrero:2021etw}. A direct computation gives
\es{Qflux}{
Q=\frac{g}{2\pi}\int_\spindle\diff A=\frac{\Delta z}{2\pi}(k_1+k_2)=\frac{\nn_1+\nn_2}{\nn_1\,\nn_2}\,,
}
which shows that only twist solutions are realized in this class. 

Interestingly, the uplift of this supersymmetric spindle solution to 11d supergravity gives a background which is $\R^{1,4}$ times the cone over a Sasaki-Einstein $Y^{p,q}$ manifold of dimension five \cite{Gauntlett:2004yd,Gauntlett:2004hh}. More on this in \cite{Ferrero:2024vmz}.

\subsubsection{Disks}\label{sec:8ddisks}

In all cases considered in the rest of this paper, multiple gauge fields are present in gauged supergravity so that it is always possible to switch off one of them while still having a solution with non-trivial gauge fields. Turning off one of the gauge fields in a spindle-type solution doubles the amount of supersymmetry that is preserved and usually gives disk-type solutions \cite{Bah:2021mzw} -- see the Introduction and references therein for more details. However, in this case only one gauge field is present, which can be turned off by setting $q=0$. One can still make choices such that the space parametrized by $(y,z)$ has the topology of a disk, with features that are completely analogous to the disk solutions introduced in \cite{Bah:2021mzw}, with the significant difference that there are no gauge fields. In particular, one finds that the correct range of $y$ is 
\es{rangedisk}{
0<y<\frac{g^2}{4}\,.
}
At $y=g^2/4$ the $z$-circle shrinks and one can choose whether to have a conical deficit or to end the space smoothly, while at $y=0$ the circle does {\it not} shrink and the space ends with a boundary where there is a curvature singularity. As discussed in \cite{Ferrero:2024vmz}, this singularity is completely cured by the uplift to 11d, where it turns out that the disk solution represents a limiting case of the spindle one, where the $Y^{p,q}$ manifold is actually the five-sphere $S^5$.

\subsection{D6-branes wrapped on a circle}\label{sec:8dcircle}

The other interesting case that one can consider is $\ell=0$ in \eqref{ansatzwrappedD6}. In this case, the condition for supersymmetry is that $s=0$, so we conclude that there is no supersymmetric solution with $\ell=0$ and non-trivial gauge fields. Even in the absence of supersymmetry, it is straightforward to see that there is no {\it compact} range of $y$ for which the solution is real and has the correct signature: for that to happen we need $P>0$, $y>0$ and $H>0$, but setting $\ell=0$ we find that $P$ is a polynomial of degree two in $y$, where the coefficient of $y^2$ is always positive. So we are forced to choose a non-compact range for $y$ and in particular $y\in(y_+,+\infty)$, where $y_-<y_+$ are the two roots of $P$, given by
\es{}{
y_\pm=-\frac{2s^2}{g^2\,q}\pm \sqrt{\frac{4s^4}{g^4\,q^2}-q}\,,
}
where to have $y_+>0$ we need $q<0$ (for $g>0$). Expanding near $y\simeq y_+$, we find that the circle parametrized by $z$ shrinks smoothly if we choose $z$ to have periodicity
\es{}{
\Delta z=\frac{16\pi\,s\,y_+}{\sqrt{-q}\,g^3\,P'(y_+)}\,,
}
while we can explore the regime $y\to +\infty$ by setting $y=1/\epsilon$ and expanding for small $\epsilon$. At leading order, we find\footnote{Note that for $\ell=0$ we can set $r=\ex^{x^5}$ and replace $\R^{1,4}\to \R^{1,5}$.}
\es{}{
\ds_8&=\epsilon^{-1/2}\,\left[\ds(\R^{1,5})+\frac{1}{g^2}\frac{\diff\epsilon^2}{\epsilon^3}+\frac{g^2}{4}\diff z^2\right]\,,\\
A&=0\,,\quad
\ex^{2\phi}=\epsilon^{-3}\,,\quad
\ex^{6\lambda}=1\,.
}
This is just the vacuum of the 8d gauged supergravity, which uplifts to the flat D6-branes in type IIA, with the only difference that one of the worldvolume directions, that parametrized by $z$, is compact. We conclude that this solutions describes the compactification on a circle of the worldvolume theory of D6-branes, in the spirit of \cite{Nunez:2023xgl} for fivebranes (with the difference that this solution is not supersymmetric).We refer the reader to \cite{Nunez:2023xgl} for more details on the field theory interpretation.

\subsection{Rotating D6-branes from black holes}\label{sec:8dBH}

It is straightforward to obtain a Wick-rotated version of \eqref{ansatzwrappedD6}, which represents an electrically charged black hole with flat horizon in eight dimensions for $\ell=0$, as described in Section \ref{sec:summary}. This is obtained by setting $s\to \ii\,s$ on top of a double analytical continuation which exchanges the role of time and that of the coordinate $z$. The resulting solution (for $\ell=0$) can be expressed as
\es{}{
\ds_8&=y^{1/6}H^{1/6}\left[-\frac{g^2\,P}{4\,H}\diff t^2+\frac{y\,\diff y^2}{g^2\,P}+\ds(\R^{6})\right]\,,\\
A&=\frac{\s}{H}\diff t\,,\quad
\ex^{2\phi}=y\,H\,,\quad
\ex^{6\lambda}=\frac{H}{y^2}\,,
}
where 
\es{}{
H=y^2+q\,,\quad 
P=H-\frac{4s^2}{g^2\,q}y\,.
}
The uplift to ten dimensions gives
\es{}{
\ds_{10}&=\frac{\sqrt{H}}{\sqrt{y}}\,\left[-\frac{g^2\,P}{4\,H}\diff t^2+\frac{y}{g^2\,P}\diff y^2+\ds(\R^6)+\frac{y}{g^2}(\diff\theta^2+\sin^2\theta\,\diff\alpha^2)\right]\,,\\
C_1&=-\cos\theta\,\diff\alpha+g\,\frac{s}{H}\,\diff t\,,\\
\ex^{2\Phi}&=\frac{H^{3/2}}{g^2\,y^{3/2}}\,,
}
and we note that as opposed to the other solutions presented in this paper the electric gauge field in 8d does not lead to a fibration of the internal space over the 8d spacetime in the IIA version of the solution. The interpretation as rotating D6-branes is therefore unclear and it would be interesting to study this background more in detail in the future.

\section{D5/NS5-branes}\label{sec:D5}

The case of D5/NS5-brane is the one that has been explored more thoroughly, mainly in the attempt to find a holographic description for 4d SYM theories, with and without flavor. Most famously, in \cite{Maldacena:2000yy} a solution of 7d $SO(4)$ gauged supergravity was presented which represents D5-branes wrapped on $S^2$ with non-abelian fluxes across the $S^2$, which is interpreted as the holographic dual to pure $\cN=1$ SYM. The analogous solution for pure $\cN=2$ SYM was presented in \cite{Gauntlett:2001ps,Bigazzi:2001aj} and it only involves abelian gauge fields. The addition of flavor to these solutions is also an interesting problem, which has been tackled using smeared branes \cite{Casero:2006pt}. However, none of these solutions, or limits thereof, falls into (the fivebrane adaptation of) our ansatz \eqref{ansatz_generic_ell}, as we will explain in Section \ref{sec:7driemann}. Solutions describing fivebranes wrapped on a circle have also appeared in the literature \cite{Nunez:2023xgl}, which arise from the uplift of the truncation of 7d supergravity considered here with the two gauge fields identified. We will generalize this to two independent gauge fields.

\subsection{7d gauged supergravity}\label{sec:7dsugra}

As we discussed in Section \ref{sec:supergravities}, for our purposes there are two interesting gaugings in 7d: the $ISO(4)$ gauging associated with type IIA NS5-branes \cite{Cvetic:2000ah} and the $SO(4)$ gauging for type IIB NS5-branes and, up to S-duality, D5-branes. All gaugings of 7d maximal supergravity were classified in \cite{Samtleben:2005bp}, but here we are interested in a simpler abelian truncation, common to the $ISO(4)$ and $SO(4)$ gaugings,  with two gauge fields $A^{(I)}$ and two scalars $\lambda_I$, for which explicit formulas were presented in \cite{Bigazzi:2001aj}. The action is
\es{action7d}{
\mathcal{S}=\frac{1}{16\pi G_N^{(7)}}\int \diff^7x \sqrt{-g}&\left[R+\mathcal{V}-5(\de_\mu \lambda_+)^2-(\de_\mu\lambda_-)^2\right.\\
&\left. -\frac{1}{4}\ex^{-4\lambda_1}(F^{(1)}_{\mu\nu})^2-\frac{1}{4}\ex^{-4\lambda_2}(F^{(2)}_{\mu\nu})^2\right]\,,
}
where $\lambda_{\pm}=\lambda_1\pm \lambda_2$, $F^{(I)}=\diff A^{(I)}$ and the scalar potential is
\es{potential}{
\mathcal{V}=4g^2\,\ex^{2(\lambda_1+\lambda_2)}\,,
}
in terms of the gauge coupling $g$. A solution to this model is said to preserve supersymmetry if the following KSE are satisfied\footnote{Note that the KSE arising from the $SO(4)$ theory have additional factors of $\gammafiveE_5$ in front of the ``superpotential'' terms in the gaugino variations. These are responsible for the different chirality of the 10d Killing spinors in the uplift to IIA and IIB. However, this does not play a significant role here so we shall neglect this subtlety for simplicity.}
\es{KSE7d}{
0=\delta\psi_\mu&=\left[\nabla_{\mu}+\frac{g}{2}(A^{(1)}_{\mu}\,\gammafiveE_{12}+A^{(2)}_\mu\, \gammafiveE_{34})+\frac{1}{2}\gammasevenL_\mu\,\slashed{\de}\lambda_+\right.\\
&\left.\quad\hspace{0.3cm}+\frac{1}{4}(\gammasevenL)^{\nu}\,\left(\ex^{-2\lambda_1}\,F^{(1)}_{\mu\nu}\,\gammafiveE_{12}+\ex^{-2\lambda_2}\,F^{(2)}_{\mu\nu}\gammafiveE_{34}\,\right)\right]\epsilon\,,\\
0=\delta \chi^{(1)}&=\left[\slashed{\de}(3\lambda_1+2\lambda_2)-g\,\ex^{2\lambda_1}+\frac{\ex^{-2\lambda_1}}{4}\slashed{F}^1\,\gammafiveE_{12}\right]\epsilon\,,\\
0=\delta \chi^{(2)}&=\left[\slashed{\de}(2\lambda_1+3\lambda_2)-g\,\ex^{2\lambda_2}+\frac{\ex^{-2\lambda_1}}{4}\slashed{F}^2\,\gammafiveE_{34}\right]\epsilon\,,
}
where $\psi_\mu$ is a gravitino and $\chi^{(I)}$ are gaugini. Note that $\epsilon$ is a spinor of both the Lorentz group $Spin(1,6)$ and the R-symmetry group $SO(5)\simeq USp(4)/\Z_2$, whose gamma matrices we denote with $\gammafiveE$. For future reference we also define
\es{AR}{
A^{(R)}\equiv A^{(1)}+A^{(2)}\,,
}
which we refer to as the R-symmetry gauge field, or the graviphoton, under which the gravitino is charged. 

Any solution of this model can be uplifted to a purely NSNS solution of either type IIA or type IIB (equivalently, so far as bosonic fields are concerned). The uplift formulas can be read from \cite{Cvetic:2000dm} and they are most conveniently expressed introducing four coordinates $\mu^i$, $i=1,\ldots,4$ subject to the constraint $\mu^i\mu^i=1$, parametrizing the embedding of $S^3$ into $\R^4$. We solve the constraint by introducing three coordinates $(\eta,\xi_1,\xi_2)$ via
\es{mui}{
\mu_1+\ii\,\mu_2=\cos\eta\,\ex^{\ii\xi_1}\,,\qquad
\mu_3+\ii\,\mu_4=\sin\eta\,\ex^{\ii\xi_2}\,,
}
with $\eta\in[0,\tfrac{\pi}{2}]$ and $\xi_i\in[0,2\pi)$. To write the uplifted solution, it is useful to introduce some notation. First, we gauge the one-forms $\diff\xi_i$ on $S^3$ with the 7d gauge fields defining
\es{Dxi}{
\Diff\xi_i\equiv \diff \xi_i-g\,A^i\,.
}
Then, we introduce the combinations
\es{DeltaU}{
\VV=\ex^{2\lambda_1}\cos^2\eta+\ex^{2\lambda_2}\sin^2\eta\,,\quad
\UU=2\left[\ex^{4\lambda_1}\cos^2\eta+\ex^{4\lambda_2}\sin^2\eta-\VV\,\left(\ex^{2\lambda_1}+\ex^{2\lambda_2}\right)\right]\,.
}
Using these, we can write the ten-dimensional string frame metric as
\es{upliftgeneralmetric}{
\ds_{10}=\ex^{2(\lambda_1+\lambda_2)}\ds_7+\frac{1}{g^2}\left(\diff\eta^2+\frac{\ex^{2\lambda_2}\cos^2\eta\,\Diff\xi_1^2+\ex^{2\lambda_1}\sin^2\eta\,\Diff\xi_2^2}{\VV}\right)\,,
}
where $\ds_7$ is the metric of the 7d solution, while the dilaton is given by
\es{upliftgeneraldilaton}{
\ex^{2\Phi}=\frac{\ex^{6(\lambda_1+\lambda_2)}}{\VV}\,,
}
and finally the NSNS three-form field strength is
\es{upliftgeneralH3}{
H_3=&\frac{\sin 2 \eta}{2g^2\,\VV^2}\left[\UU\,\diff\eta+\ex^{2(\lambda_1+\lambda_2)}\sin 2 \eta\,\diff(\lambda_1-\lambda_2)\right]\wedge\Diff\xi_1\wedge\Diff\xi_2\\
&+\frac{1}{g\,\VV}\left[\ex^{2\lambda_1}\cos^2\eta\,\Diff\xi_1\wedge\diff A_2+\ex^{2\lambda_2}\sin^2\eta\,\diff A_1\wedge\Diff\xi_2\right]\,.
}
Note that this is closed only if $F^{(1)}\wedge F^{(2)}=0$, which is necessary for consistency of the truncation. If, moreover, we also have $A^{(1)}\wedge A^{(2)}=0$, which will be the case for all solutions of interest in this paper, the associated Kalb-Ramond two-form potential can be expressed as
\es{upliftgeneralB2}{
B_2=\frac{1}{2g^2\,\VV}\left[\ex^{2\lambda_1}\cos^2\eta\,\diff\xi_1\wedge(\diff\xi_2-2g\,A^{(2)})-\ex^{2\lambda_2}\sin^2\eta\,(\diff\xi_1-2g\,A^{(1)})\wedge\diff\xi_2\right]\,.
}

Once a 7d solution is uplifted to type IIB using the formulas above, which give a Kalb-Ramond flux in ten dimensions,it can be S-dualized to a solution with three-form RR flux describing D5-branes.Given the metric $\ds_{\text{NS}}$, Kalb-Ramond three-form field strength $H_3$ and dilaton $\Phi_{\text{NS}}$ coming from (\ref{upliftgeneralmetric}-\ref{upliftgeneralB2}), the D5-brane solution is obtained replacing $H_3$ with a $F_3$ via
\es{F3_Sdual}{
\star_{\text{D}}F_3=\ex^{-2\Phi_{\text{NS}}}\star_{\text{NS}}H_3\,,
}
while the metric $\ds_{\text{D}}$ and dilaton $\Phi_{\text{D}}$ of the D5 solution are given by
\es{dsPhi_Sdual}{
\ds_{\text{D}}&=\ex^{\Phi_{\text{D}}}\ds_{\text{NS}}\,,\\
\Phi_{\text{D}}&=-\Phi_{\text{NS}}\,.
}

\subsection{The general solution}\label{sec:7dgeneral}

Here we introduce the general form of the solution that we work with in this section. We start from an ansatz\footnote{Note that here we are using a coordinate $\rho$, which plays the role of $r$ in \eqref{ansatz_generic_ell}, but appears in the solution in a different way, as typical of fivebranes \cite{Boonstra:1998mp}.}
\es{7dsol_general}{
\ds_7&=\ex^{\tfrac{4\ell\,g}{5}\rho}H^{1/5}\left[\ds(\R^{1,3})+\diff\rho^2+\frac{\diff y^2}{4g^2\,P}+\frac{P}{H}\diff z^2\right]\,,\\
A^{(I)}&=\frac{\pp_I\,\rr_I}{h_I}\diff z\,,\quad \ex^{\lambda_I+\tfrac{\ell\,g}{5}\rho}=H^{1/5}\sqrt{\frac{\pp_I}{h_I}}\,,
}
with $I=1,2$, where  
\es{7dsol_functions}{
H=h_1\,h_2\,,\quad h_I=y+\qq_I\,,
}
in terms of the seven parameters $\qq_I$, $\pp_I$ and $\rr_I$ and $\ell$. The equations of motion then require the function $P$ to satisfy
\es{7dsol_P}{
P=(\pp_1\pp_2-\ell^2)H-\frac{\rr_2^2\,h_1-\rr_1^2\,h_2}{\qq_1-\qq_2}\,.
}
Not all parameters are physical: besides the possibility of tuning $\ell$ (which leads to the discrete choice $\ell=0$ or $\ell=1$ considered below), a shift symmetry in $y$ can be used to show that the sum $\qq_1+\qq_2$ is not physical.We shall nonetheless keep both parameters explicitly in our solution for the time being, as they allow to explore special cases in a convenient way. 

\subsection{D5/NS5-branes wrapped on two-cycles}\label{sec:7dorbifolds}

Here we focus on the case $\ell=1$. We start with simpler version of \eqref{7dsol_general}, where the 2d space parametrized by $(y,z)$ is replaced by a Riemann surface with a constant curvature metric. This can always be reproduced starting from \eqref{7dsol_general} with a suitable scaling limit, in the spirit of Appendix A of \cite{Ferrero:2021etw}. We then move to the case where the solution preserves supersymmmetry and we show that it is possible to make some choices in such a way that $(y,z)$ parametrize a compact space which is either a spindle or a disk. The conditions for supersymmetry are carefully analyzed in Appendix \ref{app:7dsusy}.

\subsubsection{Warm up: Riemann surfaces}\label{sec:7driemann}

Let us consider the simpler case in which the space parametrized by $(y,z)$ is replaced with a genus-$\genus$ Riemann surfaces $\sigmag$, which we take to have a rigid metric of constant curvature, completely fixed by isometries. Our ansatz is
\es{riemannansatz}{
\ds_7&=\ex^{\tfrac{4g\,\rho}{5}}\left[\ds(\R^{1,3})+\diff\rho^2+R^2\ds(\sigmag)\right]\,,\\
F^{(I)}&=-\tfrac{1}{2}\pp_I\,\rr_I\,\vol(\sigmag)\,,\quad 
\ex^{\lambda_I+\tfrac{g\,\rho}{5}}=\sqrt{\pp_I}\,,
}
where $R$, $\pp_I$ and $\rr_I$ ($I=1,2$) are parameters that should be fixed by the equations of motion and/or supersymmetry conditions. First we observe that, opposed to AdS cases, we find that no solutions exist for $\genus>0$ -- see \cite{Paredes:2004xw} for comments on the hyperbolic case. In the genus one case, simply no solution with non-trivial gauge fields is possible, while for hyperbolic Riemann surfaces we find that the radius $R$ would be complex for real values of the charge parameters. We shall then henceforth focus on the $S^2$ case with $\genus=0$. We find that the equations of motion fix
\es{riemannparameters}{
-\rr_1=\rr_2=2R=\frac{1}{g\,\sqrt{1-\pp_1\pp_2}}\,,
}
while the constraint coming from the requirement that some amount of supersymmetry is preserved is
\es{riemannsusy}{
\pp_1+\pp_2=2\,.
}
In the case $\pp_1\in (0,1)$ and $\pp_2\in(1,2)$, this gives
\es{riemannparameters_susy}{
-\rr_1=\rr_2=2R=\frac{1}{g\,(1-\pp_1)}\,.
}
When the conditions above are met, we find that the solution preserves four Killing spinors $\epsilon$, corresponding to 4d $\cN=1$ supersymmetry on $\R^{1,3}$. Choosing a frame
\es{frameriemann}{
\ex^{a}=\ex^{\tfrac{2g\,\rho}{5}}\left(\diff x^0,\,\diff x^1,\,\diff x^2,\,\diff x^3,\,\diff \rho,\,R\,\diff \theta,\,R\,\sin\theta\,\diff\phi\right)\,,
}
we find that such spinors are subject to the projections \es{riemannprojection}{
\gammafiveE_{12}\epsilon=\gammafiveE_{34}\epsilon=\gammasevenL_{\rho\theta\phi}\epsilon\,,\quad \gammasevenL_{\rho}\epsilon=\epsilon\,.
}
The requirement \eqref{riemannsusy} is the condition for a partial topological twist, which is what allows to preserve supersymmetry, since it identifies the total R-symmetry gauge field $A^{(R)}$ defined in \eqref{AR} with the spin connection on $S^2$. Correspondingly, the projection \eqref{riemannprojection} leads to constant and chiral Killing spinors on $S^2$, see \cite{Ferrero:2021etw} for a detailed discussion. More explicitly, the total R-symmetry flux is
\es{fluxriemann}{
\charge^{(R)}\equiv\frac{g}{2\pi}\int_{S^2}\diff A^{(R)}=2=\chi(S^2)\,,
}
where $\chi(S^2)=2$ is the Euler characteristic of the sphere. We conclude by observing that this solution could not have been found by either \cite{Maldacena:2000yy} or \cite{Gauntlett:2001ps,Bigazzi:2001aj}, for the following reasons. In \cite{Maldacena:2000yy} the authors mostly consider non-abelian gauge fields, but they open with some comments on the case with $U(1)$ gauge group, corresponding to abelian minimal supergravity ({\it i.e. } only the graviphoton field $A^{(R)}$ is non-trivial). This corresponds to the limit $\pp_1=\pp_2=1$ of our solution, which however is clearly singular (see \eqref{riemannparameters_susy}).\footnote{Note that a very similar solution appears in \cite{Casero:2006pt}, where the internal space of the 10d solution is fibered on the $S^2$ using only what would be the graviphoton field from a supergravity perspective, which is precisely the case of minimal supergravity. However, we stress that \cite{Casero:2006pt} does {\it not} describe vacuum solutions and the presence of sources associated to smeared branes is crucial for their construction.} On the other hand, \cite{Gauntlett:2001ps,Bigazzi:2001aj} insist on solutions preserving eight supercharges, which in the language of \eqref{riemannansatz} could, in principle, be achieved setting either $\rr_1=0$ or $\rr_2=0$. However, once again this is not compatible with the conditions imposed by the equations of motion and supersymmetry conditions.

\subsubsection{Spindles}\label{sec:7dspindles}

Here we focus on the case $\ell=1$ of the solution \eqref{7dsol_general}. The conditions under which some supersymmetry is preserved in this case are presented in Appendix \ref{app:7dsusy}, where it is shown that in the general case with both $\rr_I\neq 0$ \eqref{7dsol_general} preserves four Killing spinors, corresponding to 4d $\cN=1$ supersymmetry on the unwrapped $\R^{1,3}$ part of the brane's worldvolume, if the parameters satisfy
\es{7dspindle_BPSconditions}{
\pp_1=\pp_2\equiv \pp\,,\quad
\rr_1-\rr_2=\qq_1-\qq_2\,.
}
We can solve the second condition by setting $\rr_I=\qq_I+c$, in terms of a new parameter $c$, which turns out to be redundant, so from now on we set
\es{}{
\rr_I=\qq_I\,.
}
The supersymmetric solution can be then written as
\es{7dspindle_susy}{
\ds_7&=\ex^{\tfrac{4\,g}{5}\rho}H^{1/5}\left[\ds(\R^{1,3})+\diff\rho^2+\frac{\diff y^2}{4g^2\,P}+\frac{P}{H}\diff z^2\right]\,,\\
A^{(I)}&=\pp\,\frac{\qq_I}{h_I}\diff z\,,\quad \ex^{\lambda_I+\tfrac{g}{5}\rho}=H^{1/5}\sqrt{\frac{\pp}{h_I}}\,,
}
where the functions are given by
\es{7dspindle_susy_functions}{
H=h_1\,h_2\,,\quad h_I=y+\qq_I\,,\quad
P=\pp^2\,H-y^2\,.
}
For the discussion of supersymmetry, it is also useful to introduce two functions $h_\pm$ as
\es{hpm_spindle}{
h_\pm=\pp\,\sqrt{H}\pm y\,,
}
in terms of which
\es{PH_hpm_spindle}{
P=h_+\,h_-\,,\quad
\sqrt{H}=\frac{h_++h_-}{2\pp}\,,\quad
y=\frac{h_+-h_-}{2}\,.
}

Let us now focus on the case in which neither of the two parameter $\qq_I$ vanishes, which corresponds to solutions with four Killing spinors. In this case, we can exploit a scaling symmetry to set $\qq_1\qq_2=1$. More precisely, we choose
\es{q12spindle}{
\qq_1=\qq_2^{-1}\equiv \qq\,.
}
In this case we can show that it is possible to make certain choices for the parameters and the range of the coordinates such that the two-dimensional space parametrized by $(y,z)$ is a spindle $\spindle\equiv \mathbb{WCP}^1_{[\nn_1,\nn_2]}$. To this end, let us define
\es{spindlemetric}{
\ds(\spindle)=\frac{\diff y^2}{4g^2\,P}+\frac{P}{H}\diff z^2\,,
}
where with the choices made so far
\es{PHspindle}{
H=(y+\qq)(y+\qq^{-1})\,,\quad 
P=\pp^2\,H-y^2\,.
}
We would like to choose a compact range for $y$, $y\in(y_1,y_2)$, such that $P$ vanishes at the two extrema of the interval (so that the circle parametrized by $z$ shrinks) and such that the solution is real in Lorentzian signature in this range. This requires $P>0$, $H>0$, $\pp/h_I>0$. Note that there is a $\Z_2$ symmetry that exchanges
\es{Z2spindle}{
y\leftrightarrow -y\,,\quad z\leftrightarrow -z\,,\quad \qq\leftrightarrow - \qq\,,\quad \pp\leftrightarrow -\pp\,,
}
therefore mapping $h_I\to -h_I$ while leaving $P$ and $H$ invariant. Since both $h_I$ must have the same sign as $\pp$ for the scalars to be real, we can use this symmetry to focus on the case $\pp>0$, $h_I>0$ without loss of generality. Moreover, note that $P$ is a degree-two polynomial in $y$, so to have $P>0$ in a compact range of values for $y$ the coefficient of $y^2$ must be negative, which only happens for $\pp<1$. The two roots of $P$ can be easily expressed in terms of the two parameters $\qq$ and $\pp$ as
\es{y12}{
y_1=\frac{\pp\left[\pp(1+\qq^2)-\sqrt{\pp^2(1-\qq^2)+4\qq^2}\right]}{2\qq(1-\pp^2)}\,,\quad
y_2=\frac{\pp\left[\pp(1+\qq^2)+\sqrt{\pp^2(1-\qq^2)+4\qq^2}\right]}{2\qq(1-\pp^2)}\,,
}
which for $\qq>0$ and $0<\pp<1$ are real and satisfy $y_1<y_2$. One can also show that in this range of parameters $y_{1,2}>-\qq$ and $y_{1,2}>-\qq^{-1}$, so that $h_I>0$ when $P>0$. We conclude that \eqref{spindlemetric} can be used to describe a Euclidean compact space when the coordinates and parameters are constrained by
\es{spindle_restrictions}{
\qq>0\,,\quad 0<\pp<1\,,\quad y\in[y_1,y_2]\,.
}

Let us now study the behavior of the metric \eqref{spindlemetric} near the poles $y=y_{1,2}$. This can be done by setting $y=y_i+g^2P'(y_i)r^2$ and expanding for small $r$, which gives
\es{spindlemetricpoles}{
\ds(\spindle)\simeq \diff r^2+\kappa_i^2\,r^2\,\diff z^2\,, \quad \kappa_i=\left|\frac{g\,P'(y_i)}{\sqrt{H(y_i)}}\right|\,.
}
We always take $g>0$, while $H>0$ is a necessary condition for the solution to be real. Moreover, the slope of $P$ is necessarily positive at $y=y_1$ and negative at $y=y_2$. So we can remove the absolute value with no ambiguities and let
\es{kappai}{
\kappa_1=+\frac{g\,P'(y_1)}{\sqrt{H(y_1)}}\,,\quad
\kappa_2=-\frac{g\,P'(y_2)}{\sqrt{H(y_2)}}\,.
}
Note that in general $\kappa_1\neq \kappa_2$, so the metric \eqref{spindlemetric} cannot be regular at both poles. Instead, we allow for conical deficits $2\pi(1-\nn_i^{-1})$ which are quantized for $\nn_{1,2}\in \N$ and for $\text{hcf}(\nn_1,\nn_2)=1$ characterize a weighted projective space $\mathbb{WCP}^1_{[\nn_1,\nn_2]}$, {\it i.e.} a spindle $\spindle$. This is realized when
\es{Deltaz_spindle}{
\Delta z=\frac{2\pi}{\kappa_1\,\nn_1}=\frac{2\pi}{\kappa_2\,\nn_2}\,.
}
Note that the (Levi-Civita) spin connection for \eqref{spindlemetric} is
\es{spinconnection}{
\omega_{\spindle}=\frac{g}{H^{3/2}}\left(P\,H'-H\,P'\right)\,\diff z\,,
}
so the Euler character of $\spindle$ can be computed using \eqref{Deltaz_spindle} as
\es{euler}{
\chi(\spindle)=\frac{1}{2\pi}\int_{\spindle}\diff\omega_{\spindle}=\left(\frac{P'(y_1)}{\sqrt{H(y_1)}}-\frac{P'(y_2)}{\sqrt{H(y_2)}}\right)\,g\,\Delta z=\frac{\nn_1+\nn_2}{\nn_1\,\nn_2}\,.
}
Since our solution contains gauge fields with magnetic flux across the spindle, such fluxes ought to be suitably quantized. On a weighted projective space, this means that we should have
\es{QIdef}{
\charge^{(I)}=\frac{g}{2\pi}\int_{\spindle}\diff  A^{(I)}=\pp\,\frac{g\,\Delta z}{2\pi}\,\left(\frac{y_1}{h_I(y_1)}-\frac{y_2}{h_I(y_2)}\right)\equiv \frac{\pflux_I}{\nn_1\,\nn_2}\,,\quad \pflux_I\in \Z\,.
}
Now, using 
\es{roots_identities}{
P'=\pp^2(h_1+h_2)-2y\,,\quad y_i=\pp\,\sqrt{h_1(y_1)h_2(y_i)}\,,
}
we can write
\es{kappainew}{
\kappa_i=2g\,\pp\,\left(1-\frac{y_i}{2}\sum_{I=1}^2\frac{1}{h_I(y_i)}\right)\,,
}
which allows us to compute the flux of the R-symmetry gauge field \eqref{AR} without solving any equation. We find
\es{QRantitwist}{
\charge^{(R)}\equiv \charge^{(1)}+\charge^{(2)}=\frac{\Delta z}{2\pi}(\kappa_2-\kappa_1)=\frac{\nn_1-\nn_2}{\nn_1\,\nn_2}\,.
}
This shows that all solutions in this class satisfy the {\it anti-twist} condition, according to the classification of \cite{Ferrero:2021etw}. The fact that only one of the two possible twists is realized should not be surprising given that there is no room for ambiguities when removing the absolute values in the definition of $\kappa_i$ in \eqref{spindlemetricpoles}, as discussed in \cite{Ferrero:2021etw}.Equation \eqref{QRantitwist} shows that the two flux parameters $\pflux_I$ are not independent, but rather constrained by the anti-twist condition
\es{ptot}{
\pflux_1+\pflux_2=\nn_1-\nn_2\,.
}

We now have all the ingredients to solve the quantization conditions \eqref{Deltaz_spindle} and \eqref{QIdef} explicitly. Some algebra shows that the parameters of the solution should be related to the integers $\nn_{1,2}$ and $\pflux_{1,2}$ by
\es{quantconditions}{
\qq=-\sqrt{\frac{\pflux_2}{\pflux_1}}\,,\quad
\pp=\frac{\sqrt{\pflux_1\pflux_2}}{\sqrt{\pflux_1\pflux_2+\nn_1\nn_2}}\,,\quad
\Delta z=\frac{2\pi}{g\,\nn_1\nn_2\,(\nn_1+\nn_2)}\frac{(\pflux_1\pflux_2+\nn_1\nn_2)^{3/2}}{\sqrt{\pflux_1\pflux_2}}\,,
}
which is valid for all values of $\nn_{1,2}\in \N$ with $\text{hcf}(\nn_1,\nn_2)=1$ and negative integer flux parameters $-\pflux_{1,2}\in \N$ constrained by \eqref{ptot}. Note that when these conditions are met, we always have $0<\pp<1$ and $\qq>0$, as required by \eqref{spindle_restrictions}. We can also rewrite the expression of the roots $y_{1,2}$ given in \eqref{y12}, as
\es{y12quant}{
y_1=-\frac{\sqrt{\pflux_1\pflux_2}}{\nn_2}\,,\qquad 
y_2=\frac{\sqrt{\pflux_1\pflux_2}}{\nn_1}\,.
}
Note that $y_1$ and $y_2$ always have opposite sign, compatibly with the fact that the solution realizes the anti-twist and the observations of \cite{Ferrero:2021etw}. We also observe that the limit $\pflux_1=\pflux_2$, which gives the abelian truncation of minimal gauged supergavity, is possible.

\subsubsection{Disks}\label{sec:7ddisks}

Let us now go back to \eqref{7dspindle_susy} and consider the special case where one of the two parameters $\qq_I$ vanishes, which preserves twice as many supercharges as the previous one (eight Killing spinors, or 4d $\cN=2$ supersymmetry of $\R^{1,3}$). Let us pick for concreteness $\qq_2=0$. In this case, we can use a scaling symmetry to set $\qq_1=1$, so the functions and parameters appearing in the supersymmetric solution in this section will be given by
\es{functions_disk}{
H=y(y+1)\,,\quad P=y\,(\pp^2-(1-\pp^2)y)\,,
}
with only one physical parameter $\pp$. Note that $P$ has again two roots, this time given by
\es{y12_disk}{
y_1=0\,,\quad y_2=\frac{\pp^2}{1-\pp^2}\,,
}
and we require $0<\pp<1$ using similar arguments to the previous section. The behavior of the metric at $y=y_2$ is similar to the spindle case, but now both $H$ and $P$ are simultaneously vanishing at $y=y_1=0$. Let us consider the 7d metric near this point: setting $y=g^2\,\pp^2\,r^2$ and expanding for small $r$ we find (rescaling $z\to\pp^{-1}\,z$)
\es{diskboundarysing}{
\ds_7\simeq r^{2/5}\,\left[\ds(\R^{1,3})+\diff\rho^2+\diff r^2+\diff z^2\right]\,,
}
which has a curvature singularity for $r=0$, but the circle parametrized by $z$ does not shrink at that point. Hence, we must terminate the metric with a boundary at $y=y_1=0$. This is precisely the situation first described in \cite{Bah:2021mzw} and following that we require that the space parametrized by $(y,z)$ is a disk $\disk$, with a boundary (and a curvature singularity) at $y=y_1=0$ and potentially a conical singularity at $y=y_2$. We then set
\es{diskmetric}{
\ds(\disk)=\frac{\diff y^2}{4g^2\,P}+\frac{P}{H}\diff z^2\,,
}
and study the behavior near $y_2$ by setting $y=y_2-g^2\,\pp^2\,r^2$ and expanding for small $r$. We find
\es{diskmetricpole}{
\ds(\disk)\simeq \diff r^2+\kappa^2\,r^2\,\diff z^2\,,\quad \kappa=g\,\pp\,(1-\pp^2)\,.
}
Requiring that the metric has a quantized conical deficit $2\pi(1-\nn^{-1})$ at this point gives the condition
\es{Deltaz_disk}{
\Delta z=\frac{2\pi}{\kappa\,\nn}\,,
}
which can be used to compute the Euler character
\es{euler_disk}{
\chi(\disk)=\frac{1}{\nn}\,.
} 
The flux of the non-vanishing gauge field $A^{(1)}$ should also be quantized. Given the presence of a boundary, to do so we can choose a gauge where $A^{(1)}$ is regular at $y=y_2$, namely we define
\es{Aregular}{
\widetilde{A}^{(1)}=A^{(1)}-\pp(1-\pp^2)\diff z\,, \quad 
\left. \widetilde{A}^{(1)}\right|_{y=y_2}=0\,.
}
Then, the flux of $A^{(1)}$ equals minus the holonomy of the gauge field across the boundary
\es{holonomy}{
\hol\equiv\text{hol}_{\de \disk}\left(\widetilde{A}^{(1)}\right)=\frac{g}{2\pi}\oint_{y=y_2}\widetilde{A}^{(1)}=-\frac{g}{2\pi}\int_{\disk}\diff \widetilde{A}^{(1)}=\frac{g\,\Delta z}{2\pi}\pp^3\,.
}
We can solve the two conditions \eqref{Deltaz_disk} and \eqref{holonomy} for the parameters $\pp,\Delta z$ in terms of the physical data $\nn$ and $\hol$, which should characterize the dual field theory. We find
\es{parquant_disk}{
\pp=\frac{\sqrt{\hol\,\nn}}{\sqrt{1+\hol\,\nn}}\,,\quad
\Delta z=\frac{2\pi}{g\,\nn}\frac{(1+\hol\,\nn)^{3/2}}{\sqrt{\hol\,\nn}}\,.
}

\subsection{D5/NS5-branes wrapped on a circle}\label{sec:7dcircle}

Let us now move to the case $\ell=0$, in which we generalize the solutions of \cite{Nunez:2023xgl} describing fivebranes wrapped on a circle to a case with two independent gauge fields. Differently from the $\ell=1$ case discussed above, we find that generic supersymmetric solutions preserve eight Killing spinors. The conditions for supersymmetry are
\es{BPS_circle}{
\pp_1=\pp_2=\pp\,,\quad \rr_1=\rr_2=\rr\,,
}
which implies that one cannot switch off only one of the two gauge fields: they are either both vanishing or both non-vanishing. We can get rid of unphysical parameters as follows. First, note that the shift $y\to y+b$, $\qq_I\to \qq_I-b$ is a symmetry of the solution, which we can use it to fix
\es{qfix_circle}{
\qq_1=-\qq_2=\qq\,.
}
Moreover, a scaling symmetry can be used to fix the ratio $\rr/\pp$ to an arbitrary value, so long as $\rr\neq 0$: we use it to set $\rr=1$ and end up again with a solution that has two independent physical parameters, $\qq$ and $\pp$.\footnote{When $\rr=0$ both gauge fields vanish and sixteen supercharges are preserved. This is a domain-wall solution and in particular it can be shown that it is equivalent, after the uplift to type II, to the solution of \cite{Sfetsos:1998xd} describing a uniform distribution of NS5-branes on a circle in a double scaling limit.}

With these choices, the supersymmetric solution describing fivebranes wrapped on a circle can be written as
\es{circlesol_susy}{
\ds_7&=H^{1/5}\,\left[\ds(\R^{1,4})+\frac{\diff y^2}{4g^2\,P}+\frac{P}{H}\diff z^2\right]\,,\\
A^{(I)}&=\frac{\pp}{h_I}\diff z\,,\quad
\ex^{\lambda_I}=H^{1/5}\,\sqrt{\frac{\pp}{h_I}}\,,
}
where the functions are
\es{functions_circle}{
H=h_1\,h_2\,,\quad h_1=y+\qq\,,\quad h_2=y-\qq\,,\quad 
P=\pp^2\,H-1\,.
}
Note that we have incorporated the $\rho$ direction in $\R^{1,4}$, as discussed in Section \ref{sec:philosophy}. Similarly to Section \ref{sec:7dspindles}, we find it useful to introduce two functions
\es{hpm_circle}{
h_\pm=\pp\,\sqrt{H}\pm 1\,,
}
in terms of which
\es{PH_hpm_circle}{
P=h_+\,h_-\,,\quad \sqrt{H}=\frac{h_++h_-}{2\pp}\,,\quad
1=\frac{h_+-h_-}{2}\,.
}

So far as global considerations on the solution \eqref{circlesol_susy} are concerned, we mainly refer the reader to the analysis of \cite{Nunez:2023xgl}. Here we limit to observe that one may wonder whether the space parametrized by $(y,z)$ can be made compact as in the case of spindle/disk solutions, but as it turns out this is not the case. This is because 
\es{Pcircle}{
P=\pp^2\,y^2-(1+\pp^2\qq^2)\,,
}
and since $\pp$ is real $P$ is always convex, hence there is no compact range where $P$ is positive. Take $\pp>0$ (which can be always achieved using the $\Z_2$ symmetry discussed in the previous section), which requires $h_I>0$. Then the roots of $P$ are
\es{rootsPcircle}{
y_\pm=\pm\frac{\sqrt{1+\pp^2\qq^2}}{\pp}\,,
} 
with $y_-<0<y_+$. Note that the solution is also symmetric under $\qq\to -\qq$ so we can take $\qq>0$ for definiteness and to have $P>0$ we need to take $y\in[y_+,+\infty)$. Then, $h_I>0$ requires $y_+>\qq$, but this is always satisfied for $\pp>0$ and $\qq>0$.

We can then study the behavior of the solution near the two extrema of the range of $y$. For $y=y_+$ the circle parametrized by $z$ shrinks smoothly if we require that the periodicity of $z$ is
\es{Deltaz_circle}{
\Delta z=\frac{\pi}{g}\frac{1}{\pp^2\sqrt{1+\pp^2\qq^2}}\,.
}
The behavior near $y=+\infty$ can be studied by setting $y=1/x$ and expanding for small $x$. We find
\es{circle_infinity}{
\ds_7\simeq x^{-2/5}\,\left[\ds(\R^{1,4})+\frac{\diff x^2}{4g^2\,\pp^2\,x^2}+\pp^2\,\diff z^2\right]\,,
}
which is a singular behavior. Changing $x\to \ex^{-s}$ (with $s\to +\infty$) one can see that the singularity is of the linear dilaton type, with $s$ providing the linear dilaton direction.

Let us also note that the solutions discussed in Section \ref{sec:7dspindles} in the case $\ell=1$ have a similar global structure to those discussed here, in a certain regime of parameters. In particular, to obtain a compact range for $y$ in Section \ref{sec:7dspindles} we had to require that $|\pp|<1$, but if $|\pp|\ge 1$ we obtain non-compact solutions in that case too. 

\subsection{Rotating D5/NS5-branes from black holes}\label{sec:7dBH}

Let us now consider a double analytical continuation of \eqref{7dsol_general}, which leads to the solution
\es{7dsol_general_Wick}{
\ds_7&=\ex^{\tfrac{4\ell\,g}{5}\rho}H^{1/5}\left[-\frac{P}{H}\diff t^2+\frac{\diff y^2}{4g^2\,P}+\ds(\R^{1,3})+\diff\rho^2\right]\,,\\
A^{(I)}&=\frac{\pp_I\,\rr_I}{h_I}\diff t\,,\quad \ex^{\lambda_I+\tfrac{\ell\,g}{5}\rho}=H^{1/5}\sqrt{\frac{\pp_I}{h_I}}\,,
}
with $I=1,2$, where
\es{7dsol_functions_Wick}{
H=h_1\,h_2\,,\quad h_I=y+\qq_I\,,
}
as before. The equations of motion then fix the function $P$ to
\es{7dsol_P_Wick}{
P=(\pp_1\pp_2-\ell^2)H+\frac{\rr_2^2\,h_1-\rr_1^2\,h_2}{\qq_1-\qq_2}\,.
}
Note that, so long as the functions appearing in the solution are concerned, the only difference with \eqref{7dsol_general}, is a sign in the expression of $P$, which comes from mapping $\rr_I\to \ii\,\rr_I$, necessary to compensate for $\diff z \to -\ii\,\diff t$. 

Up to some trivial rescalings, the uplifted version of the solution (in the string frame) is
\es{uplift_generalsol_Wick}{
\ds_{10}&=
\pp_1\pp_2\,\left[-\frac{P}{H}\diff t^2+\frac{\diff y^2}{4g^2\,P}\right]+\ds(\R^4)+\diff\rho^2+\frac{1}{g^2}\ds(Y_3)\,,\\
\ds(Y_3)&=\diff\eta^2+\frac{\pp_2\,h_1\,\cos^2\eta}{\Delta}\left(\diff\xi_1-g\,\frac{\pp_1\rr_1}{h_1}\diff t\right)^2+\frac{\pp_1\,h_2\,\sin^2\eta}{\Delta}\left(\diff\xi_2-g\,\frac{\pp_2\rr_2}{h_2}\diff t\right)^2\,,\\
B_2&=\frac{1}{2g^2\,\Delta}\left[\pp_1\,h_2\,\cos^2\eta\,\diff\xi_1\wedge\left(\diff\xi_2-2g\,\frac{\pp_2\rr_2}{h_2}\diff t\right)\right.\\
&\hspace{2cm} \left.-\pp_2\,h_1\,\sin^2\eta\,\left(\diff\xi_1-2g\,\frac{\pp_1\rr_1}{h_1}\diff t\right)\wedge\diff\xi_2\right]\,,\\
\Phi&=-\frac{g\,\ell}{\sqrt{\pp_1\pp_2}}\,\rho-\frac{1}{2}\log\Delta\,,
}
where we have introduced the combination
\es{Deltadef_Wick}{
\Delta=\pp_1\,h_2\,\cos^2\eta+\pp_2\,h_1\,\sin^2\eta\,.
}

The 10d solution \eqref{uplift_generalsol_Wick}, in the case $\ell=0$, is closely reminiscent of the solutions presented in Section 4 of \cite{Sfetsos:1999pq} (see also \cite{Nakayama:2008kg}), describing rotating NS5-branes. In fact those solutions arise as a special case of \eqref{uplift_generalsol_Wick}, as we can show explicitly with a suitable change of coordinates and choice of parameters. More precisely, if we set\footnote{Note that the coordinate $r$ below is the coordinate $\rho$ of \cite{Sfetsos:1999pq,Nakayama:2008kg}, while since we are studying the case with $\ell=0$ we can set our $\rho\to x^5$ and replace $\ds(\R^4)+\diff\rho^2\to \ds(\R^5)$.}
\es{comparetoBH}{
&y\to r^2+a_1^2-\qq_1\,,\quad \eta\to \theta-\frac{\pi}{2}\,,\quad
\xi_I\to \phi_I\,,\\
&\qq_2\to a_2^2-a_1^2+\qq_1\,,\quad \pp_I\to 1\,,\quad
\rr_I\to a_I\,,\quad g\to 1\,.
}
However, note that this solution is {\it not} supersymmetric. Also, note that our solution \eqref{uplift_generalsol_Wick} contains more physical parameters and is therefore more general.

So far as supersymmetry is concerned, we find that the only supersymmetric solutions with $\ell=1$ have vanishing gauge fields and reduce to domain walls with sixteen supercharges. On the other hand, for $\ell=0$ we do find supersymmetric solutions with non-vanishing gauge fields, which preserve eight supercharges. They can be found setting 
\es{susywickell0}{
\pp_1=\pp_2\equiv \pp\,,\quad \rr_1=\rr_2\equiv \rr\,,
}
and they correspond to the Wick rotation of the supersymmetric solutions \eqref{circlesol_susy}. Using the same logic as in that section, we can set $\qq_1=-\qq_2\equiv \qq$ and $\rr=1$. The resulting function $P$ is
\es{Psusywickell0}{
P=\pp^2(y-y_+)(y-y_-)\,,\quad y_\pm=\pm \frac{\sqrt{\pp^2\qq^2-1}}{\pp}\,.
}
Similarly to the discussion around \eqref{circlesol_susy} we can focus on $\pp,\qq>0$ and $y>\qq$,but since we always have $y_+<\qq$ we are forced to take $y\in (\qq,+\infty)$. This is problematic because $y=y_+$ is the would-be-horizon of the would-be-black hole, while $y=\qq$ is a curvature singularity which is necessarily a naked singularity in the supersymmetric case.

\section{D4-branes}\label{sec:D4}

While the case of D4-branes wrapped on a spindle/disk in the presence of D8 flavor branes has already been discussed in \cite{Suh:2021aik,Faedo:2021nub,Giri:2021xta}, there is no solution available in the literature that describes the case of D4-branes alone. Here, in the spirit of what we did in the previous sections, we consider solutions of this type, where only the four-form RR flux is present in type IIA. There is actually a straightforward way of obtaining D4-branes solutions from existing M5-branes solutions, which involves a (rather trivial) reduction of the latter along one of the worldvolume directions, as we shall soon review.

\subsection{6d gauged supergravity}\label{sec:6dsugra}

The relevant theory here is the the 6d maximal $SO(5)$ gauged supergravity discussed in \cite{Cowdall:1998rs}. We focus on a truncation analogous to that of \cite{Ferrero:2021wvk} where two gauge fields $A^{(I)}$ ($I=1,2$) are present, with field strengths $F^{(I)}=\diff A^{(I)}$. In addition to the two scalars $\lambda_I$ appearing in \cite{Ferrero:2021wvk},  we consider an additional dynamical scalar $\sigma$ which plays an analogous role to the dilaton appearing in the reduction from 11d supergravity to type IIA. The action for this model is
\es{6daction}{
S=\frac{1}{16\pi\,G_N^{(6)}}\int \diff^6x\,\sqrt{-g}\,&\left[R+\mathcal{V}-5(\de\lambda_+)^2-(\de \lambda_-)^2-80(\de\sigma)^2\right.\\
&\quad \left. -\frac{1}{4}\ex^{-4(\lambda_1+\sigma)}\left(F^{(1)}_{\mu\nu}\right)^2-\frac{1}{4}\ex^{-4(\lambda_2+\sigma)}\left(F^{(2)}_{\mu\nu}\right)^2\right]\,,
}
where $\lambda_\pm=\lambda_1\pm \lambda_2$ and the scalar potential is
\es{V6d}{
\mathcal{V}=\frac{g^2}{2}\ex^{4\sigma}\left(8\ex^{2(\lambda_1+\lambda_2)}-4\ex^{-2(2\lambda_1+\lambda_2)} -4\ex^{-2(\lambda_1+2\lambda_2)}+ \ex^{-8(\lambda_1+\lambda_2)}\right)\,.
}
A solution of this model preserves supersymmetry if the following KSE are satisfied 
\es{KSE6d}{
0&=\delta\psi_{\mu}=\Big[\nabla_\mu-\frac{g}{2}(A^{(1)}\gammafiveE_{12}+A^{(2)}\gammafiveE_{34})-\frac{5\,g}{16}\ex^{2\sigma-4(\lambda_1+\lambda_2)}\gammasixL_\mu-\frac{5}{8}\slashed{\partial}\lambda_+\,\gamma_\mu \\
&\hspace{2cm}+\frac{1}{4}\ex^{-2\sigma}\left(\gammasixL\right)^{\nu}\left(\ex^{-2\lambda_1}\,F^{(1)}_{\mu\nu}\,\gammafiveE_{12}+\ex^{-2\lambda_2}\,F^{(2)}_{\mu\nu}\,\gammafiveE_{34}\right)\Big]\epsilon\,,\\
0&=\delta\chi^{(0)}=\Big[\slashed{\de}(8\sigma-\lambda_1-\lambda_2)+\frac{g}{2}\ex^{2\sigma-4\lambda_1-4\lambda_2}\Big]\epsilon\,,\\
0&=\delta\chi^{(1)}=\Big[\slashed{\partial}(3\lambda_1+2\lambda_2)+g\,\ex^{2\sigma}\left(\ex^{2\lambda_1}-\ex^{-4(\lambda_1+\lambda_2)}\right)+\frac{1}{4}\ex^{-2(\lambda_1+\sigma)}\slashed{F}^{(1)}\,\gammafiveE_{12}\Big]\epsilon\,,\\
0&=\delta\chi^{(2)}=\Big[\slashed{\partial}(2\lambda_1+3\lambda_2)+g\,\ex^{2\sigma}\left(\ex^{2\lambda_2}-\ex^{-4(\lambda_1+\lambda_2)}\right)+\frac{1}{4}\ex^{-2(\lambda_2+\sigma)}\slashed{F}^{(2)}\,\gammafiveE_{34}\Big]\epsilon\,.
}
where $\psi_{\mu}$ is the gravitino, while $\chi^{(0)}$ is a fermion in the same multiplet as the scalar $\sigma$, while $\chi^{(I)}$, $I=1,2$, are in the same multiplet as $\lambda^{(I)}$, from the point of view of minimal supersymmetry. As usual, $\epsilon$ is a supersymmetry parameter which in this case is both a spinor of $Spin(1,5)$ and of the R-symmetry group $SO(5)\simeq USp(4)/\Z_2$, whose gamma matrices we denote with $\gammafiveE$.

Let us also briefly discuss this model can be obtained from the circle reduction of the one considered in \cite{Ferrero:2021wvk}. Let $x$ be the direction along which we reduce and denote the 7d fields with a hat: $\hat{g}_{\hat{\mu}\hat{\nu}}$, $\hat{A}^{(I)}$ and $\hat{\lambda}_I$, where $\hat{\mu}=(\mu,x)$, $\mu=0,\ldots,5$. We assume that $\partial_x$ is a Killing vector and moreover
\es{reductionconditions}{
\hat{g}_{\hat{\mu}x}=0\,,\quad
\hat{A}^{(I)}_x=0\,.
}
Then the 7d fields reduce as
\es{reduction7dto6d}{
\ds_7=\ex^{4\sigma}\,\ds_6+\ex^{-16\sigma}\,\diff x^2\,,\quad
\hat{A}^{(I)}=A^{(I)}\,,\quad
\hat{\lambda}_I=\lambda_I\,.
}
Viceversa, given any of the 6d solutions that we discuss below, it can be uplifted to 7d using the formulas above. Then, 7d solutions can be uplifted to 11d on $S^4$ using the formulas given in \cite{Ferrero:2021wvk}. Finally, reducing from 11d supergravity to type IIA along $x$ gives a solution in IIA corresponding to the uplift of solutions to the model \eqref{6daction} on $S^4$. Since this is quite standard and explicit formulas are available elsewhere for all these uplifts/reductions, we shall not present them explicitly here in the general case.

\subsection{The general solution}\label{sec:6dgeneral}

Following the same ideas as the previous sections, we make the ansatz
\es{6dgeneral}{
\ds_6&=\frac{y^{1/4}H^{1/4}}{r^{5\ell/2}}\,\left[\ds(\R^{1,2})+r^{2(\ell-1)}\diff r^2+r^{2\ell}\,\left(\frac{y}{g^2\,P}\diff y^2+\frac{g^2\,P}{4\,H}\diff z^2\right)\right]\,,\\
A^{(I)}&=\frac{s_I}{h_I}\,\diff z\,,\quad
\ex^{2\lambda_I}=p_I\,\frac{y^{2/5}\,H^{2/5}}{h_I}\,,\quad
\ex^{8\sigma}=\frac{p_0\,r^{\ell}}{y^{1/10}\,H^{1/10}}\,,
}
where the functions are
\es{functions6dgeneral}{
H=h_1\,h_2\,,\quad h_I=y^2+q_I\,,
}
and we find that the equations of motion fix
\es{Petc6d}{
P=\frac{q_2\,s_1^2-q_1\,s_2^2}{q_1q_2(q_1-q_2)\ell^2}H+\frac{4\ell^2\,y}{(q_2\,s_1^2-q_1\,s_2^2)g^2}(q_1\,s_2^2\,h_1-q_2\,s_1^2\,h_2)\,,
}
as well as
\es{p012}{
p_0=\frac{q_2\,s_1^2-q_1\,s_2^2}{q_1q_2(q_1-q_2)\ell^2}\,,\quad p_1=p_2=1\,.
}
As in the other cases, our parametrization is redundant. In the case $\ell=1$ we can use a scaling symmetry to show that the product $s_1\,s_2$ is not physical, so we can introduce a new parameter $s$ and write $s_I=s\,q_I$. In terms of this, we find \es{fixscaling6d}{
p_0=s^2\,,\quad p_I=1\,,\quad
s_I=s\,q_I\,.
}

The case $\ell=0$ is a bit more subtle because naively there is a singularity in \eqref{Petc6d} and \eqref{p012} for $\ell=0$. However, it turns out that solutions still exist in this case and they can be obtained by setting
\es{}{
q_2\,s_1^2-q_1\,s_2^2=\tilde{p}_0\,\ell^2\,,
}
and then taking the limit $\ell\to 0$. After the limit, we can use a scaling symmetry to set $\tilde{p}_0=1$. Moreover, we still have to solve the constraint $q_2\,s_1^2-q_1\,s_2^2=0$, which we can do by setting
\es{}{
s_I=s\,\sqrt{q_I}\,,
}
in terms of a new parameter $s$.

\subsection{D4-branes wrapped on compact cycles}\label{sec:6dorbifolds}

\subsubsection{Warm up: Riemann surfaces}\label{sec:6driemann}

As for the other cases, let us present first the solution corresponding to branes wrapped on a Riemann surface $\sigmag$. We make the ansatz
\es{}{
\ds_6&=\frac{H_0^2}{r^{5/2}}\,\left[\ds(\R^{1,2})+\diff r^2+r^2\,R^2\,\ds(\sigmag)\right]\,,\\
F^{(I)}&=\diff A^{(I)}=s_I\,\vol(\sigmag)\,,\quad
\ex^{2\lambda_I}=p_I\,,\quad
\ex^{8\sigma}=p_0\,r\,,
}
in terms of unknown parameters $H_0$, $R$, $s_I$, $p_0$ and $p_I$. We find that supersymmetric solutions exist for all genera and values of the free parameters always exist such that these solutions are real and in Lorentzian signature. In all cases, we find that $H_0$ and $s_{1,2}$ can be expressed in terms of the other parameters as
\es{}{
H_0=\frac{2p_1^2p_2^2}{g\,p_0^{1/4}}\,,\quad
s_1=\frac{8R^2\,p_1^2p_2^2}{g}(p_1^3p_2^2-1)p_1\,,\quad
s_2=\frac{8R^2\,p_1^2p_2^2}{g}(p_1^2p_3^2-1)p_2\,,
}
where only the combination $H_0\,p_0^{1/4}$ is physical, so we could set {\it e.g.} $p_0=1$. For $\genus=1$ the radius $R$ is redundant and we could set $R=1$. For $\genus=0$ ($\kappa=1$) and $\genus>1$ ($\kappa=-1$), instead, we have
\es{RkappaD4}{
R=\frac{1}{2\sqrt{2}p_1p_2\sqrt{\kappa\,[p_1^2\,p_2^2\,(p_1^2+p_2^2)-p_1-p_2]}}\,,
}
while in all cases the parameters $p_1$ and $p_2$ are constrained by
\es{}{
2p_1^2p_2^2(p_1+p_2)=3\,.
}
In the torus case ($\genus=1$, $\kappa=0$), moreover, the combination appearing in the denominator of \eqref{RkappaD4} also vanishes, so $p_1$ and $p_2$ are completely constrained and they satisfy
\es{}{
p_1=\left(\frac{33+19\sqrt{3}}{4}\right)^{1/5}\,,\quad
p_2=\left(\frac{33-19\sqrt{3}}{4}\right)^{1/5}\,,
}
or equivalently a version of the equation above with $p_1$ and $p_2$ exchanged. Note that for all cases we have
\es{}{
s_1+s_2=\frac{\kappa}{g}\,,
}
which is the topological twist condition.

\subsubsection{Spindles and disks}\label{sec:6dspindles}

Let us focus on the case $\ell=1$, which is the only one possible as discussed earlier in this section. We find that the solution \eqref{6dgeneral} preserves superymmetry when
\es{6dsusy}{
s=1\,,
}
in which case there are four Killing spinors. The supersymmetric solution reads\footnote{Note that we have chosen a gauge such that the Killing spinors are uncharged under $\partial_z$.}
\es{6dsusyspindle}{
\ds_6&=\frac{y^{1/4}H^{1/4}}{r^{5/2}}\,\left[\ds(\R^{1,2})+\diff r^2+r^2\,\left(\frac{y}{g^2\,P}\diff y^2+\frac{g^2\,P}{4\,H}\diff z^2\right)\right]\,,\\
A^{(I)}&=\left(\frac{q_I}{h_I}-\frac{1}{4}\right)\,\diff z\,,\quad 
\ex^{2\lambda_I}=\frac{y^{2/5}\,H^{2/5}}{h_I}\,,\quad
\ex^{8\sigma}=\frac{r}{y^{1/10}\,H^{1/10}}\,,
}
where the functions are
\es{functions6d}{
H=h_1\,h_2\,,\quad P=H-\frac{4}{g^2}y^3\,,\quad h_I=y^2+q_I\,.
}
Using \eqref{reduction7dto6d}, it can be shown that this matches exactly the solution discussed in \cite{Ferrero:2021wvk} (which is obtained setting $g=1$). The global analysis for the spindle case can be found in \cite{Ferrero:2021wvk,Ferrero:2021etw}, while for the disk case it was presented in \cite{Bah:2021mzw,Bah:2021hei} and we shall not repeat it here. We remind the reader that the analysis of \cite{Ferrero:2021wvk} shows that only twist solutions are realized in this case, and moreover the case of identical gauge fields does not give rise to spindle solutions.

\subsubsection{Bonus: four-dimensional orbifolds}\label{sec:6d4dorbifolds}

We note that an interesting solution has been obtained in \cite{Cheung:2021igt} describing M5-branes wrapped on four-dimensional orbifolds, which are the fibration of a spindle over either i) a Riemann surface or ii) another spindle. More general solutions were recently constructed in \cite{Martelli:2023oqk}, which involve M5-branes wrapped on so-called quadrilaterals. A trivial reduction along one of the worldvolume directions, along the lines of \eqref{reduction7dto6d}, generalizes those M5-brane solutions in 11d to solutions with D4-branes in type IIA, and correspondingly solutions of 7d $SO(5)$ gauged supergravity to solutions of 6d $SO(5)$ gauged supergravity (generally in a larger truncation than the one considered here).

\subsection{D4-branes wrapped on a circle}

As we discussed previously in this section, the solution with $\ell=0$ can be obtained by taking a suitable limit of the general case \eqref{6dgeneral}. Fixing some redundancy in the parameterization, we obtain
\es{6dcircle}{
\ds_6&=y^{1/4}H^{1/4}\,\left[\ds(\R^{1,3})+\left(\frac{y}{g^2\,P}\diff y^2+\frac{g^2\,P}{4\,H}\diff z^2\right)\right]\,,\\
A^{(I)}&=s\,\frac{\sqrt{q_I}}{h_I}\,\diff z\,,\quad
\ex^{2\lambda_I}=\frac{y^{2/5}\,H^{2/5}}{h_I}\,,\quad
\ex^{8\sigma}=\frac{1}{y^{1/10}\,H^{1/10}}\,,
}
where
\es{}{
H=h_1\,h_2\,,\quad
P=H+\frac{4s^2}{g^2}y\,,\quad h_I=y^2+q_I\,,
}
in terms of the three parameters are $q_{1,2}$ and $s$. It turns out that there is no choice of parameters for which the solutions is supersymmetric while also having non-trivial gauge fields. 

Note that we need $y>0$ for the solution to be real, and for $q_I>0$ the function $P$ never vanishes since it is the sum of two positive terms. Hence, we should take $q_I\to-\tilde{q}_I$ and $s\to \ii\,\tilde{s}$, with $\tilde{q}_I>0$. The solution is still real, and now 
\es{}{
P=H-\frac{4\tilde{s}^2}{g^2}y\,,
}
can vanish for $y>0$. The explicit expression of the roots is complicated, but one can use Descartes' rule of signs to obtain some information on their behavior. $P$ is a polynomial of degree four, and it has either 0 or 2 positive roots and either 0 or 2 negative roots. Hence, we cannot find a compact range for $y$ with $y>0$ and $P>0$. Instead, like in the other cases with $\ell=0$ we can take $y\in(y_+,+\infty)$, where $y_+$ is the largest root of $P$, assuming that the parameters are chosen such that $y_+>0$. The circle parametrized by $z$ shrinks smoothly at $y=y_+$ provided that one makes a suitable choice for the periodicity of $z$. Moreover, expanding near $y\sim \infty$ we find that the solution reduces to the domain wall describing the near-horizon region of D4-branes, but with one compact direction (parametrized by $z$). We conclude that the solution describes holographically a situation in which the worldvolume theory of D4-branes is placed on a circle.

\subsection{Rotating D4-branes from black holes}\label{sec:6dBH}

A double analytical continuation of \eqref{6dcircle}, where the role of time and that of the coordinate $z$ are swapped, as well as taking $s\to \ii\,s$, gives a new solution which describes electrically charged black holes in 6d, with a flat horizon. Much like spindles and disks, they can be found from the dimensional reduction of a 7d solution: the AdS$_7$ black holes of \cite{Cvetic:1999xp}. In particular, in the case of flat horizon in 7d one can reduce along one of the horizon directions and obtain directly a Wick-rotated version of \eqref{6dcircle}. We have checked that no supersymmetry is preserved when the gauge fields are non-trivial. Moreover, the uplift of the 7d black holes in \cite{Cvetic:1999xp} to 11d gives a solution with rotating M5-branes, whose dimensional reduction along one of the worldvolume directions corresponds to the 10d uplift of the Wick-rotated version of \eqref{6dcircle} and describes rotating D4-branes in type IIA.

\section{D2-branes}\label{sec:D2}

The case of wrapped D2-branes has not been considered in the literature before, to the best of our knowledge, to the point that the details of the abelian truncation that we consider here were not previously available in the literature. 

\subsection{4d gauged supergravity}\label{sec:4dsugra}

The relevant supergravity for D2-branes is a 4d $ISO(7)$ gauged theory which can be obtained as a deformation of the maximal $SO(8)$ gauging \cite{Hull:1984yy}. Here we are interested in a truncation with three abelian gauge fields $A^{(I)}$, $I=1,2,3$, (the Cartans of the compact subgroup $SO(7)\subset ISO(7)$) and the four parity-even scalars $\phi_i$ ($i=0,1,2,3$) of the $E_{7(7)}/SU(8)$ coset that are neutral under such gauge fields. While the $ISO(7)$ theory and its magnetic deformation (which encodes the Romans mass in type IIA) has been thoroughly investigated in a series of papers \cite{Guarino:2015jca,Guarino:2015qaa,Guarino:2015vca}, we did not find the details of this truncation in the literature, in particular so far as the supersymmetry variations are concerned. We refer the reader to Appendix \ref{app:4dsugraderivation} for some comments on the derivation of the Lagrangian and KSEs of the model that we consider in this section, which relies on the classification of all gaugings of 4d maximal supergravity \cite{deWit:2007kvg}. Here we limit to present the action
\es{4daction}{
S=\frac{1}{16\pi G^{(4)}_N}\int \diff^4 x \sqrt{-g}\left[R+\mathcal{V}-\frac{1}{2}\sum_{i=0}^3(\partial\phi_i)^2-\frac{1}{4}\sum_{I=1}^3(X^{(I)})^{-2}(F^{(I)}_{\mu\nu})^2\right]\,,
}
where we have introduced the three combinations of scalars
\es{}{
X^{(1)}=\ex^{\tfrac{1}{2}(\phi_1-\phi_2-\phi_3)}\,,\quad
X^{(2)}=\ex^{\tfrac{1}{2}(-\phi_1+\phi_2-\phi_3)}\,,\quad
X^{(3)}=\ex^{\tfrac{1}{2}(-\phi_1-\phi_2+\phi_3)}\,,
}
and we also define
\es{}{
X^{(0)}=\frac{1}{2}\ex^{\tfrac{1}{2}(\phi_1+\phi_2+\phi_3)-\phi_0}\,.
}
In terms of these auxiliary objects, the scalar potential is given by
\es{}{
\mathcal{V}=g^2\,\left[\sum_{0\le i<j\le 3}X^{(i)}X^{(j)}-\frac{1}{2}(X^{(0)})^2\right]\,.
}
The conditions for supersymmetry to be preserved by solutions of \eqref{4daction} can be expressed as 
\es{KSE4d}{
0=\delta \psi_\mu&=\left[\nabla_\mu-\frac{g}{4}\left(A^{(1)}_\mu \,\gammaeightE_{12}+A^{(2)}_\mu \,\gammaeightE_{34}+A^{(3)}_\mu \,\gammaeightE_{56}\right)+\frac{g}{8}\sum_{i=0}^3 X^{(i)}\,\gammafourL_\mu \right.\\
& \left.\hspace{1cm}+\frac{i}{16}\left((X^{(1)})^{-1}\,\slashed{F}^{(1)}\,\gammaeightE_{12}+(X^{(2)})^{-1}\,\slashed{F}^{(2)}\,\gammaeightE_{34}+(X^{(3)})^{-1}\,\slashed{F}^{(3)}\,\gammaeightE_{56}\right)\,\gammafourL_\mu \right]\,\epsilon\,,\\
0=\delta\chi^{(0)}&=\left[\slashed{\partial}\phi_0+g\,X^{(0)}\right]\,\epsilon\,,\\
0=\delta\chi^{(1)}&=\left[\slashed{\partial}\phi_1-\frac{g}{2}\left(X^{(0)}+X^{(1)}-X^{(2)}-X^{(3)}\right)\right.\\
& \left.\hspace{1cm}-\frac{i}{4}\left((X^{(1)})^{-1}\,\slashed{F}^{(1)}\,\gammaeightE_{12}-(X^{(2)})^{-1}\,\slashed{F}^{(2)}\,\gammaeightE_{34}-(X^{(3)})^{-1}\,\slashed{F}^{(3)}\,\gammaeightE_{56}\right)\right]\,\epsilon\,,\\
0=\delta\chi^{(2)}&=\left[\slashed{\partial}\phi_1-\frac{g}{2}\left(X^{(0)}-X^{(1)}+X^{(2)}-X^{(3)}\right)\right.\\
& \left.\hspace{1cm}-\frac{i}{4}\left(-(X^{(1)})^{-1}\,\slashed{F}^{(1)}\,\gammaeightE_{12}+(X^{(2)})^{-1}\,\slashed{F}^{(2)}\,\gammaeightE_{34}-(X^{(3)})^{-1}\,\slashed{F}^{(3)}\,\gammaeightE_{56}\right)\right]\,\epsilon\,,\\
0=\delta\chi^{(3)}&=\left[\slashed{\partial}\phi_1-\frac{g}{2}\left(X^{(0)}-X^{(1)}-X^{(2)}+X^{(3)}\right)\right.\\
& \left.\hspace{1cm}-\frac{i}{4}\left(-(X^{(1)})^{-1}\,\slashed{F}^{(1)}\,\gammaeightE_{12}-(X^{(2)})^{-1}\,\slashed{F}^{(2)}\,\gammaeightE_{34}+(X^{(3)})^{-1}\,\slashed{F}^{(3)}\,\gammaeightE_{56}\right)\right]\,\epsilon\,,
}
where $\epsilon$ is a $Spin(1,3)$ spinor which is also a fundamental the $SU(8)$ R-symmetry group, while $\gammaeightE$ are gamma matrices for $SO(8)\subset SU(8)$. From the point of view of 4d $\cN=2$ supersymmetry, $\phi_0$ and $\chi^{(0)}$ belong to a hypermultiplet (containing additional scalars that we have set to zero) and $\phi_0$ uplifts to the dilaton in type IIA, while the vectors arise from the graviphoton of the gravity multiplet as well as two vector multiplets. Two of the three scalars $\phi_I$ ($I=1,2,3$) belong to said two vector multiplets, while the third comes from a third vector multiplet whose vector has been set to zero (it is a RR field coming from the reduction of the RR one-form potential of type IIA).

The uplift formulas for this model can be found in \cite{Guarino:2015vca} (see also \cite{Guarino:2015qaa} for additional details and \cite{Guarino:2015jca} for a streamlined presentation). We give an explanation on how to use those formulas in Appendix \ref{app:4dsugraderivation}, but we will not present explicit expressions here since they are rather cumbersome.

\subsection{The general solution}\label{sec:4dgeneral}

Inspired from the previous sections as well as some guess-work, we formulate the ansatz
\es{4dsol_general}{
\ds_4&=\frac{(y\,H)^{1/2}}{r^{7\ell/3}}\,\left[-\diff t^2+r^{2(\ell-1)}\diff r^2+r^{2\ell}\,\left(\frac{4y\,\diff y^2}{g^2\,P}+\frac{9g^2\,P}{16\,H}\diff z^2\right)\right]\,,\\
A^{(I)}&=\frac{s_I}{h_I}\diff z\,,\quad (I=1,2,3)\,,\\
\ex^{\phi_0}&=\frac{y}{p_0\,r^{2\ell/3}}\,,\quad
\ex^{\phi_I}=\frac{\sqrt{H}}{p_I\,\sqrt{y}\,h_I\,r^{\ell/3}}\,,\quad (I=1,2,3)\,,
}
where the functions $h_I$ and $H$ are given by
\es{}{
H=\prod_{I=1}^3h_I\,,\quad h_I=y^2+q_I\,,
}
and $p_0$, $p_I$, $q_I$, $s_I$ are parameters. The equations of motion constrain the parameters via
\es{s_D2}{
\sqrt{p_0}=p_1=p_2=p_3\equiv p\,,\quad
s_I=\sqrt{p\,q_I\,(q_I\,\ell^2+s^2)}\,,
}
in terms of two new parameters $p$ and $s$, while also fixing the function $P$ to be
\es{}{
P=p\,H-\frac{16}{9g^2}y(\ell^2\,y^2-s^2)\,.
}
Moreover, it is simple to show that a scaling symmetry exists that allows to fix $p=1$ so we will work with this choice from now on.

\subsection{D2-branes wrapped on two-cycles}\label{sec:4dorbifolds}

\subsubsection{Warm up: Riemann surfaces}\label{sec:4driemann}

As for the other branes, let us begin by discussing the case of D2 wrapped on Riemann surfaces. To look for solutions of this type, we make the ansatz
\es{D2sigmag}{
\ds_4&=\frac{H_0^2}{r^{7/3}}\,\left[-\diff t^2+\diff r^2+r^2\,R^2\,\ds(\sigmag)\right]\,,\\
F^{(I)}&=s_I\,\vol(\sigmag)\,,\\
\ex^{\phi_0}&=\frac{p_0}{r^{2/3}}\,,\quad
\ex^{\phi_I}=\frac{p_I}{r^{1/3}}\,,
}
and in this case we find that supersymmetric solutions exist for all genera. In all cases, the equations of motion and supersymmetry fix
\es{}{
H_0=\frac{2\sqrt{p_1p_2p_3}}{(p_1+p_2+p_3)g}\,,\quad
p_0=\frac{3p_1p_2p_3}{2(p_1+p_2+p_3)}\,,
}
while to express the remaining conditions it is convenient to introduce the following combination of parameters
\es{}{
\Lambda=(p_1+p_2+p_3)^2-6(p_1p_2+p_1p_3+p_2p_3)\,.
}
In terms of the curvature $\kappa$, with $\kappa=+1$ for $\genus=0$, $\kappa=0$ for $\genus=1$ and $\kappa=-1$ for $\genus>0$, we find that reality of the solution requires that
\es{}{
\text{sgn}(\Lambda)=\kappa\,,
}
which in particular implies $\Lambda=0$ for the torus case $\genus=1$. In all cases we find values of the parameters $p_I$ for which this condition is satisfied, showing that real solutions exist for all genera. In the torus case the parameter $R$ is redundant and we just set it to 1. In the other cases, we find
\es{}{
R=\sqrt{\frac{3}{2}}\frac{p_1+p_2+p_3}{\sqrt{|\Lambda|}}\,.
}
For non-zero curvature, we find that the charge parameters $s_I$ are fixed by
\es{}{
s_I=\frac{2\kappa}{g\,\Lambda}\,p_I(3p_1-2(p_1+p_2+p_3))\,,
}
while for the torus
\es{}{
s_I=\frac{3}{3g(p_1+p_2+p_3)^2}(3p_I-2(p_1+p_2+p_3))\,.
}
In all cases we have
\es{}{
s_1+s_2+s_3=-\frac{2}{g}\kappa\,,
}
which is the topological twist condition. A generic solution preserves two supercharges, and this number doubles whenever one of the $s_I$ is switched off.

\subsubsection{Spindles}\label{sec:4dspindles}

Let us now consider \eqref{4dsol_general} in the case with $\ell=1$, and recall that we have used the scaling symmetry to set $p=1$. The constraint given by the supersymmetry conditions simply reads
\es{}{
s=0\,,
}
so that at the end of the day we can write the supersymmetric solution as
\es{4dsusyspindle}{
\ds_4&=\frac{(y\,H)^{1/2}}{r^{7/3}}\,\left[-\diff t^2+\diff r^2+r^{2}\,\left(\frac{4y\,\diff y^2}{g^2\,P}+\frac{9g^2\,P}{16\,H}\diff z^2\right)\right]\,,\\
A^{(I)}&=\frac{q_I}{h_I}\diff z\,,\quad
\ex^{\phi_0}=\frac{y}{r^{2/3}}\,,\quad
\ex^{\phi_I}=\frac{\sqrt{H}}{\sqrt{y}\,h_I\,r^{1/3}}\,,
}
where the functions are given by
\es{}{
H=\prod_{I=1}^3h_I\,,\quad h_I=y^2+q_I\,,\quad
P=H-\frac{16}{9g^2}y^3\,,
}
in terms of the three physical parameters $q_I$ ($I=1,2,3$). We now show that a choice of range for the coordinates and values of the parameters exists, such that
\es{}{
\ds(\spindle)=\frac{4y\,\diff y^2}{g^2\,P}+\frac{9g^2\,P}{16\,H}\diff z^2\,,
}
describes the metric on a spindle $\spindle$. First, note that we need $y>0$, $P>0$ and $h_I>0$ (the latter implies $H>0$) for the solution to be real and we want to be able to take $y\in(y_1,y_2)$ such that the reality conditions are met in this range, with $y_{1,2}$ roots of $P$. Expanding the metric near such roots we find 
\es{}{
\ds(\spindle)\simeq \diff\rho^2+\kappa_i^2\rho^2\diff z^2\,,
}
and using $P(y_i)=0$ as well as $P(y_1)>0$ and $P(y_2)<0$ we find
\es{}{
\kappa_1=\frac{9g^3}{64y_1^2}P'(y_1)\,,\quad
\kappa_2=-\frac{9g^3}{64y_1^2}P'(y_2)\,,
}
and since there is no ambiguity in the choice of sign we expect that only one twist should exist. The quantization of the conical deficits at the poles requires, as usual,
\es{deltazkappai_D2}{
\Delta z=\frac{2\pi}{\kappa_1\,\nn_1}=\frac{2\pi}{\kappa_2\,\nn_2}\,,
}
for two coprime integers $\nn_1$ and $\nn_2$. We note that using
\es{}{
P'=2y\left(h_1h_2+h_1h_3+h_2h_3-\frac{8y}{3g^2}\right)\,,\quad
y_i^3=\frac{9g^2}{16}H(y_i)\,,
}
we can rewrite them as
\es{kappai_D2}{
\kappa_1=-\frac{3g}{4}+\frac{g}{2}y_1^2\sum_{I=1}^3\frac{1}{h_I(y_1)}\,,\quad 
\kappa_2=\frac{3g}{4}-\frac{g}{2}y_2^2\sum_{I=1}^3\frac{1}{h_I(y_2)}\,,
}
which we can use to compute the R-symmetry flux without solving the regularity conditions explicitly. In particular, we quantize the fluxes via
\es{}{
Q^{(I)}=\frac{g}{4\pi}\int_{\spindle}\diff A^{(I)}=\frac{g\,\Delta z}{4\pi}\left(\frac{y_1^2}{h_I(y_1)}-\frac{y_2^2}{h_I(y_2)}\right)\equiv \frac{\pflux_I}{\nn_1\,\nn_2}\,,
}
for some integers $\pflux_I$, and we find that \eqref{deltazkappai_D2} and \eqref{kappai_D2} can be combined to obtain, for the total (R-symmetry) flux,
\es{}{
Q^{(R)}=\sum_{I=1}^3Q^{(I)}=\frac{\nn_1+\nn_2}{\nn_1\,\nn_2}\,,
}
showing that these solutions always realize the twist condition. Note that this is expected because we need $y_{1,2}>0$ for the solution to be real, which should indeed correspond to the twist according to the observations of \cite{Ferrero:2021etw}. We will not attempt to solve the regularity conditions explicitly here, rather we limit to provide numerical values of the parameter $q_I$ such that the necessary conditions for the reality of the solution are all met. Setting $g=4/3$ to simplify the expression of $P$, we find that for 
\es{}{
q_1=\frac{3}{10}\,,\quad
q_2=\frac{3}{4}\,,\quad
q_3=-\frac{1}{100}\,,
}
we have $y_1\simeq 0.18$ and $y_2\simeq 0.23$. One can check that for $y\in(y_1,y_2)$ we have $h_I>0$ for all $I=1,2,3$ and $P>0$, so all conditions are met. The case which would be the simplest to analyze, in which all gauge fields are identified ($q_1=q_2=q_3$), corresponding to minimal gauged supergravity, does not contain spindle solutions since in that case one cannot choose a suitable compact range for $y$ where all necessary conditions are met.

\subsubsection{Disks}\label{sec:4ddisks}

Let us now go back to \eqref{4dsusyspindle} and set one of the charge parameters to zero: for definiteness, let us choose $q_3=0$. This choice doubles the supersymmetry and we now have a solution with four supercharges. In this case, $P$ has a root at $y=0$ so one has to choose a range $y\in(y_1,y_2)$ with $y_1=0$ and $y_2>0$. The analysis of the regularity conditions is similar to the other cases so we will not repeat it here, since once again the regularity conditions are hard to solve explicitly. However, we can show explicitly that there is at least one choice of parameters which allows to choose a disk topology for the space parametrized by $(y,z)$. Setting $g=4/3$ for simplicity, we find that when 
\es{}{
q_1=\frac{3}{10}\,,\quad
q_2=\frac{7}{10}\,,\quad
q_3=0\,,
}
the polynomial $P$ has three real roots $0\le y_1<y_2<y_3$: a double root at $y_1=0$ and two simple roots $y_2\simeq 0.33$, $y_3\simeq 0.43$, with $P>0$ for $y\in (y_1,y_2)$. This choice meets all necessary criteria.

\subsection{D2-branes wrapped on a circle}\label{sec:4dcircle}

While solutions of the form \eqref{4dsol_general} certainly exist with $\ell=0$, depending on the three parameters $q_I$ as well as the parameter $s$ introduced in \eqref{s_D2}, we find that supersymmetry requires that $s=0$, which in this case sets all gauge fields to zero. We conclude that supersymmetric solutions with $\ell=0$ are simply domain walls with sixteen supercharges, which are not interesting for our purposes. We have then obtained a four-parameter family of non-supersymmetric solutions given by
\es{4dcircle}{
\ds_4&=(y\,H)^{1/2}\,\left[\ds(\R^{1,1})+\frac{4y\,\diff y^2}{g^2\,P}+\frac{9g^2\,P}{16H}\diff z^2\right]\,,\\
A^{(I)}&=s\,\frac{\sqrt{q_I}}{h_I}\diff z\,,\quad
\ex^{\phi_0}=y\,,\quad
\ex^{\phi_I}=\frac{\sqrt{H}}{\sqrt{y}}\frac{1}{h_I}\,,
}
where the functions are
\es{}{
H=\prod_{I=1}^3h_I\,,\quad
P=H-\frac{16s^2}{9g^2}y\,,\quad
h_I=y^2+q_I\,,
}
in terms of the paramters $s$ and $q_I$. Reality of the solution requires $y>0$ and $q_I>0$, which imply that the polynomial $P$ has no negative roots and at most two positive roots. Hence, there is no range of $y$ to make the space parametrized by $(y,z)$ compact, as in other cases with $\ell=0$. Rather, we consider $y\in (y_+,+\infty)$, where $y_+>0$ is the largest root of $P$ (assuming a choice of parameters such that $y_+$ is real). This can be used to describe a dual 3d theory compactified on a circle, in the spirit of the previous sections.

\subsection{Rotating D2-branes from black holes}\label{sec:4dBH}

As in the other cases, it is possible to perform a double analytical continuation of the solution \eqref{4dsol_general}, by setting $\diff z\to \ii \,\diff t$ and $s_I\to -\ii\,s_I$, which effectively changes the sign inside the square root in \eqref{s_D2}. Since a choice of parameters is still possible such that all the quantities are real, we conclude that we do have interesting solutions after Wick rotation. On the other hand, once we also request supersymmetry we find that this requires again that $s=0$. For $\ell=1$ we then end up with imaginary gauge fields and the solution should be discarded, while for $\ell=0$ the gauge fields vanish in the BPS limit $s=0$, so we conclude that there are no interesting supersymmetric solutions after Wick rotation. On the other hand, for $\ell=0$ we have electrically charged black holes with charge parameters $q_I$ and mass parameter $s$, which uplift to rotating D2-branes in ten dimensions, even though they are not supersymmetric.

\section{Discussion}\label{sec:discussion}

In this paper we have considered the consistent truncations of type II supergravity associated with the near-horizon region of D$p$-branes, with $p=2,4,5,6$, as well as NS5-branes in the case $p=5$. These are the so-called ``non-conformal'' branes, since their near-horizon region is not AdS and the associated worldvolume theories are not conformal. Despite the lower amount of symmetry with respect to the case of M2-, D3- and M5-branes, which have received significantly more attention due to their role in the AdS/CFT correspondence, we find that several interesting solutions can be found for non-conformal branes which are completely analogous to existing solutions for their conformal relatives, a fact that can be explained by the observations of \cite{Boonstra:1998mp}. Specifically, we have considered three types of solutions, as summarized in Section \ref{sec:summary}: i) branes wrapped on compact two-cycles: either Riemann surfaces or spindles \cite{Ferrero:2020laf} or disks \cite{Bah:2021mzw}, ii) branes wrapped on a circle \cite{Anabalon:2021tua} and iii) rotating branes \cite{Cvetic:1999xp}.

In this paper we have focused almost exclusively on presenting and studying new solutions of supergravity, with little attention to the dual field theories. A major difficulty in this sense is the lack of conformal invariance: only few solutions of this type have been considered in the literature and it would certainly be interesting to investigate the non-gravitational side of the holographic duality. In this sense, we would like to point out the work \cite{Bobev:2018ugk}, where solutions representing spherical D$p$-branes are obtained from the gauged supergravities that we consider here, whose on-shell action matches the free energy of SYM theory on $S^{p+1}$, computed with localization. In terms of wrapped branes, it is interesting to observe that other solutions describing fivebranes wrapped on $S^2$ have been previously considered in the literature \cite{Maldacena:2000yy,Gauntlett:2001ps,Bigazzi:2001aj}, in the study of the holographic dual of 4d SYM with four or eight supercharges. It would be interesting to investigate the relation between those solutions and the ones presented here for branes wrapped on Riemann surfaces, as well as the analogous situation in other dimensions. The case of spindles and disks is even more mysterious, since the dual theories are not particularly well-understood even in cases with conformal invariance\footnote{See \cite{Inglese:2023wky,Inglese:2023tyc} for recent progress in the study of supersymmetric localization for field theories on spindles.}. It is possible that some progress can be made for disks, due to the higher amount of supersymmetry, which was instrumental to the analysis of \cite{Bah:2021mzw}. On the other hand, the field theory dual to branes wrapped on a circle has been investigated thoroughly for the case of fivebranes in \cite{Nunez:2023xgl}, in a special case of the more general solutions presented here. It should be possible to carry out a similar analysis in other dimensions, although only fivebranes preserve supersymmetry for this type of solutions -- it would also be interesting to investigate why this is the case.

As discussed in Section \ref{sec:summary}, we expect the solutions presented here for branes wrapped on two-cycles to be the IR limit of certain ``full-flow'' solutions, much like the case of extremal black holes in AdS$_4$, but with the crucial difference that no AdS emerges either in the UV or in the IR. Moreover, it is already clear that our ansatz does not capture the most general IR behavior for non-conformal branes wrapped on a two-cycle: as an example, this is not the case for the interesting fivebrane solutions presented in \cite{Maldacena:2000yy,Gauntlett:2001ps,Bigazzi:2001aj}. It is quite possible therefore that allowing a more general dependence on $r$ could unveil more interesting physics, in all dimensions, leading to less singular solutions. In the case of D6-branes, indeed, a solution for D6-branes wrapped on $S^2$ with ``frozen'' dependence on $r$ gives the conifold in 11d, while a more general $r$ dependence yields the small resolution of the conifold, which is certainly less singular. Similar comments are possible for spindle solutions \cite{Ferrero:2024vmz}. Similarly, it should be possible to consider non-abelian gauge fields like in \cite{Maldacena:2000yy}, which could also lead to more interesting and less singular physics.

In the study of supersymmetric AdS solutions describing branes wrapped on compact cycles, interesting geometrical structures have emerged from the classification of $\ads_2\times Y_9$ solutions in 11d and $\ads_3\times Y_7$ solutions in type IIB, which has led to the discovery of Gauntlett-Kim (GK) geometries \cite{Kim:2005ez,Kim:2006qu,Gauntlett:2007ts}, are also relevant for spindles. The study of GK geometries has proven instrumental for the understanding of the dual field theories from the supergravity perspective, with tools such as the dual of $c-$ and $\mathcal{I}$-extremization \cite{Gauntlett:2018dpc,Couzens:2018wnk,Gauntlett:2019roi} and more recently gravitational blocks \cite{Faedo:2022rqx,Boido:2022iye,Boido:2022mbe}. Moreover, results from GK geometry and analogous classification problems were recently reformulated in terms of equivariant localization \cite{Martelli:2023oqk,Colombo:2023fhu,BenettiGenolini:2023kxp,BenettiGenolini:2023yfe,BenettiGenolini:2023ndb,BenettiGenolini:2024kyy}. An interesting problem is that of extending these results to the case of non-conformal branes, where as in the ansatz \eqref{ansatz_generic_ell1} the AdS factor in the metric would be replaced by a metric which is AdS only after a Weyl rescaling by a suitable power of the dilaton, in the spirit of \cite{Boonstra:1998mp}. 

Focusing momentarily of spindle solutions, we have provided several new examples which could help to shed light on some of the less clear aspects of the construction, such as the geometric interpretation of anti-twist solutions, or what controls which type of twist can be realized. See Section \ref{sec:summary} for some comments on this, where a pattern seems to emerge relating the allowed twists and the dimensional of the internal space in the uplift from gauged supergravity. It would also be interesting to extend the solutions found here to branes wrapped on four-dimensional orbifolds \cite{Cheung:2022ilc,Bomans:2023ouw,Couzens:2022lvg,Faedo:2024upq,Macpherson:2024frt}.

The same local solutions (up to analytical continuation) used to discuss spindles and disks can be also employed to study the holographic duals of Rényi entropies \cite{Huang:2014gca,Nishioka:2014mwa,Bianchi:2016xvf,Hosseini:2019and}, so it would be interesting to investigate whether a different global completion of our metrics could serve a similar purpose. Still in the spirit of generalizing what can be done with AdS spindle solutions, it is quite likely that a consistent truncation on the spindle is possible, in the spirit of \cite{Cheung:2022ilc}, which should generalize the consistent truncation for NS5-branes wrapped on $S^2$ considered in \cite{Cheung:2021igt}.

\section*{Acknowledgments}

The authors would like to thank Jerome Gauntlett, Carlos Nunez, Dario Martelli, Leonardo Rastelli, Martin Ro\v{c}ek and James Sparks for useful discussions and comments on the draft. P.F. is extremely grateful to Jerome Gauntlett, Dario Martelli and James Sparks for several interesting conversations on this topic.

\newpage

\appendix

\section{Gamma matrices}\label{app:gammas}

In this paper we use the symbol $\gamma$ to denote spacetime gamma matrices, {\it i.e.} satisfying a Clifford algebra in flat Lorentzian space with mostly plus signature. We use a superscript to denote the dimensionality of the underlying Lorentzian spacetime. In a similar spirit, we use the symbol $\tau$ to denote gamma matrices satisfying the Euclidean Clifford algebra, also indicating the corresponding dimension with a superscript. In both cases, for even dimensions the chirality matrix is denoted with a subscript $\star$. Contractions of differential forms with Clifford algebra elements are performed with unit weight, that is for a $p$-form $\omega_{(p)}$ we define
\es{}{
\slashed{\omega}=\omega_{\mu_1\ldots\mu_p}\gamma^{\mu_1\ldots \mu_p}\,.
}

\section{Supersymmetry conditions in 7d gauged supergravity}\label{app:7dsusy}

In this Appendix we discuss the supersymmetry conditions associated with the 7d solution \eqref{7dsol_general} and exhibit explicit Killing spinors. We distinguish the cases $\ell=1$ and $\ell=0$, showing that supersymmetry is generally doubled in the latter. While we express our formulas in a covariant way as much as possible, it is useful to introduce a specific representation for the gamma matrices in order to present the explicit spinors.Given the form of the ansatz \eqref{7dsol_general}, we find it useful to write the 7d gamma matrices $\gammasevenL_\mu$ in a 7=4+1+2 split, where the 4 represent four $\R^{1,3}$ directions, the 1 is the $\rho$ direction (which we is going to take to be a chirality gamma matrix from the perspective of $\R^{1,3}$), while the 2 represents spindle directions. We then write
\es{gamma7dsplit_spindle}{
\gammasevenL_{\alpha}=\gammafour_{\alpha}\otimes \gammatwoE_{\star}\,,\quad
\gammasevenL_{\rho}=\gammafour_\star\otimes \gammatwoE_\star\,,\quad
\gammasevenL_y=\mathbb{1}_4\otimes \gammatwoE_1\,,\quad
\gammasevenL_z=\mathbb{1}_4\otimes \gammatwoE_2\,,
}
with $\alpha=0,\ldots,3$ denoting the $\R^{1,3}$ directions. A convenient representation for the 4d gamma matrices in terms of Pauli matrices $\sigma^i$, $i=1,2,3$, is
\es{gamma4d}{
\gammafour_0=\ii\,\sigma^1\otimes \mathbb{1}_2\,,\quad
\gammafour_1=\sigma^2\otimes \sigma^1\,,\quad
\gammafour_2=\sigma^2\otimes \sigma^2\,,\quad
\gammafour_3=\sigma^2\otimes \sigma^3\,,
}
with chirality matrix
\es{4dchirality}{
\gammafour_\star=\ii\prod_{\alpha=0}^3\gammafour_\alpha=\sigma^3\otimes \mathbb{1}_2\,,
}
while for the 2d part we pick
\es{gamma2d}{
\gammatwoE_1=\sigma^1\,,\quad
\gammatwoE_2=\sigma^2\,,\quad
\gammatwoE_\star=-\ii \gammatwo_1\gammatwo_2=\sigma^3\,.
}
Moreover, a convenient choice of $SO(5)\simeq USp(4)/\Z_2$ gamma matrices $\gammafiveE_A$, $A=1,\ldots,5$, is given by
\es{gammaflavor}{
\gammafiveE_1&=\sigma^1\otimes \sigma^3\,,\quad
\gammafiveE_2=\sigma^2\otimes \sigma^3\,,\quad
\gammafiveE_3=\sigma^3\otimes \sigma^3\,,\\
\gammafiveE_4&=\mathbb{1}_2\otimes \sigma^1\,,\quad
\gammafiveE_5=\mathbb{1}_2\otimes \sigma^2\,.
}

\subsection{$\ell=1$: spindles and disks}

Let us consider the solution \eqref{7dspindle_susy}. We use the frame 
\es{spindle7dframeconformal}{
\ex^{a}=\Omega\,{\ext}^{a}\,,\qquad a=0,\ldots,6\,,\qquad
\Omega=\ex^{\tfrac{2g\,\rho}{5}}H^{1/10}\,,
}
where
\es{spindle7dframe}{
\ext^{\alpha}=\diff x^{\alpha}\,,\quad (\alpha=0,\ldots,3)\,,\quad
\ext^{\rho}=\diff\rho\,,\quad
\ext^{y}=\frac{\diff y}{2g\,\sqrt{P}}\,,\quad
\ext^{z}=\frac{\sqrt{P}}{\sqrt{H}}\diff z\,,
}
and $x^{\alpha}$ are coordinates on $\R^{1,3}$. With this choice, we find that the four Killing spinors $\epsilon$ preserved by \eqref{7dspindle_susy} are subject to the projections
\es{spindle7dprojections}{
\gammafiveE_{12}\epsilon=\gammafiveE_{34}\epsilon=\gammasevenL_{\rho y z}\epsilon\,,\quad \mathcal{M}\epsilon+\epsilon=0\,,
}
where the projector $\mathcal{M}$ is 
\es{Pspindle7d}{
\mathcal{M}=\sin\theta\,\gammasevenL_y+\cos\theta\,\gammasevenL_{\rho}\,,\quad 
\sin \theta=\frac{\sqrt{P}}{\pp\,\sqrt{H}}=\frac{2\sqrt{h_+h_-}}{h_++h_-}\,,\quad \cos\theta=\frac{y}{\pp\,\sqrt{H}}=\frac{h_+-h_-}{h_++h_-}\,.
}
We note that the limit in which \eqref{7dspindle_susy} reproduces the sphere solution \eqref{riemannansatz} is a limit in which $P=0$ (hence $y^2=\pp\,H$): in that limit $\theta=0$ and the projection \eqref{Pspindle7d} reduces to \eqref{riemannprojection}. The matrix $\mathcal{M}$ is the analogue of the projectors first introduced in \cite{Ferrero:2021etw}.

These are the conditions found from the gaugino equations, while the gravitino equation allows for a direct determination of the Killing spinors. To express them, we need to introduce some auxiliary objects. First, let us define two vectors transforming in the spinorial representation of the R-symmetry group $USp(4)$, $u_{\pm}$, which satisfy\footnote{Using the explicit gamma matrices \eqref{gammaflavor}, we have
\es{upmexplicit}{
u_+=(0,1,\ii,0)\,,\qquad
u_-=(1,0,0,-\ii)\,.
}
}
\es{upm}{
\gammafiveE_{12}u_{\pm}=\gammafiveE_{34}u_{\pm}=\pm \ii\, u_{\pm}\,.
}
Note that there is a unique solution for $u_\pm$ (up to normalization). Then, we introduce four-dimensional chiral spinors $\chi_{\pm}$ satisfying 
\es{chipm}{
\gammafour_\star\chi_{\pm}=\pm\chi_{\pm}\,,
}
where $\gammafour_\star$ is the chirality matrix in 4d and note that there are two such spinors for each choice of sign. Finally, we introduce the 2d spinor
\es{spindlespinor}{
\zeta=
\sqrt{\Omega}
\begin{pmatrix}
-\sin\tfrac{\theta}{2}\\
\cos\tfrac{\theta}{2}
\end{pmatrix}\,,
}
where
\es{alpha2}{
\sin\tfrac{\theta}{2}=\frac{\sqrt{h_-}}{\sqrt{h_++h_-}}\,,\qquad \cos\tfrac{\theta}{2}=\frac{\sqrt{h_+}}{\sqrt{h_++h_-}}\,.
}
Note that we have introduced the 2d charge conjugate $\zeta^c=B\,\zeta^{\star}$, where $B$ is such that $B\,(\gammatwo_{1,2})^{*}B^{-1}=\gammatwo_{1,2}$, with $\gammatwo_{1,2}$ 2d gamma matrices. Note that $\zeta$ satisfies
\es{projspindle2d}{
M\,\zeta+\zeta=0\,,\quad
M=\sin\theta\,\gammatwo_1-\ii\,\cos\theta\gammatwo_{12}\,,
}
where $\gammatwo_1$ is along $\diff y$ and $\gammatwo_2$ along $\diff z$. 

In terms of these building blocks, we can write the four Killing spinors $\epsilon$ as
\es{spinors7dspindle}{
\epsilon=u_+\otimes \chi_+\otimes \zeta+u_-\otimes \chi_-\otimes \zeta^c\,,
}
where $u_{\pm}$ take care of the flavor structure while $\chi_{\pm}\otimes \zeta^{(c)}$ are 7d spacetime spinors, and the four degrees of freedom can be thought of as arising from $\chi_{\pm}$ (two each). 

As discussed in the previous subsection, the cases $\qq_1=0$ or $\qq_2=0$ correspond to a supersymmetry enhancement and the solution has eight Killing spinors. This is because one of the gauge fields decouples from the Killing spinor equations, so one has to impose only half of the projections in \eqref{upm}: either only $\gammafiveE_{12}$ or only $\gammafiveE_{34}$. The spinors then still have the form \eqref{spinors7dspindle}, but each $u_\pm$ now carries two degrees of freedom rather than one.

Let us conclude the discussion of Killing spinors with some observation about how supersymmetry is preserved in spindle solutions. For what follows we will need to perform a frame rotation, so we introduce Lorentz generators in the vector representation
\es{Lorentz7dvecrtor}{
(\mathcal{J}^{(\mathrm{v})}_{ab})_{c}^{\,\,\,d}=\eta_{ac}\delta_{b}^{d}-\eta_{bc}\delta_{a}^{d}\,,
}
as well as Lorentz generators in the spinorial representation
\es{Lorentz7dspinor}{
\mathcal{J}^{(\mathrm{s})}_{ab}=\frac{1}{2}\gammasevenL_{ab}\,,
}
normalized in such a way as to satisfy the same algebra. We then observe that if we rotate the Killing spinors \eqref{spinors7dspindle} with
\es{rotatespinors}{
\epsilon'\equiv \ex^{{\theta}\,\mathcal{J}^{(\mathrm{s})}_{\rho y}}\epsilon\,,
}
the rotated spinors can be written as
\es{rotatedspinors}{
\epsilon'=\sqrt{\Omega}\,\left[u_+\otimes \chi_+\otimes \xi_-+u_-\otimes \chi_-\otimes \xi_+\right]\,,
}
where $\xi_\pm$ are constant, chiral spinors on the spindle satisfying
\es{chiralspindlespinors}{
\gammatwo_{\star}\xi_{\pm}=\pm \xi_{\pm}\,,
}
and $\xi_-=B\,(\xi_+)^*$. Moreover, the rotated 7d spinors $\epsilon'$ satisfy the projection $\gammasevenL_\rho\epsilon'+\epsilon'=0$, and since the rotation does not affect the flavor structure nor $\ex^\rho \wedge \ex^y$ we conclude that the rotated spinors satisfy
\es{rotatedproj}{
\gammafiveE_{12}\epsilon'=\gammafiveE_{34}\epsilon'=\gammasevenL_{\rho y z}\epsilon'\,,\quad \gammasevenL_{\rho}\epsilon'+\epsilon'=0\,,
}
which are exactly the same projections as in the $S^2$ case, see eq.  \eqref{riemannprojection}. Thus, we conclude that the spinors have exactly the same structure as in solutions with topological twist, but in a frame which is not necessarily the ``obvious'' one. In particular, the rotated frame
\es{rotationframe}{
(\ex^{b})'=\ex^{{\theta}\,(\mathcal{J}^{(\mathrm{v})}_{\rho y})_a^{\,\,\,b}}\ex^{a}\,,
}
satisfies, after extracting a conformal factor as in \eqref{spindle7dframeconformal},
\es{w12}{
\ext^\rho=\cos\tfrac{\theta}{2}\diff w^1\,,\qquad
\ext^y=\sin\tfrac{\theta}{2}\diff w^2\,,
}
where $w^{1,2}$ are new coordinates which are fixed in terms of $(y,\rho)$ via
\es{w12fix}{
w^1&=f_1(y)+\rho\,,\qquad\qquad w^2=f_2(y)-\rho\,, 
}
where the functions $f_{1,2}$ satisfy
\es{f12}{
f_1'&=\frac{\tan\theta}{2g\,\sqrt{P}}\,,\quad 
f_2'=\frac{\cot\theta}{2g\,\sqrt{P}}\,.
}

We conclude by making a comment on the analogue of what we just discussed in AdS spindle solutions. In that case, the Killing spinors provide the fermionic generators of a superconformal algebra, so one can always group them in super-translation generators and super-partners of special conformal transformations. The logic above applies to the first group of spinors, whose existence is not related the supersymmetry enhancement typical of AdS solutions, where the role of $\rho$ here is played by the AdS radial coordinate.

\subsection{$\ell=0$: circles}

As we discussed, the solution for $\ell=0$ preserves eight Killing spinors when the gauge fields are non-vanishing. To discuss the structure of these spinors,  we introduce a frame
\es{framewarpedcircle}{
\ex^a=\Omega\,\ext^a\,,\quad a=0,\ldots,6\,,\quad
\Omega=H^{1/10}\,,
}
where
\es{framecircle}{
\ext^{\bar\alpha}=\diff x^{\bar\alpha}\,\quad ({\bar \alpha}=0,\ldots,4)\,,\quad
\ext^y=\frac{\diff y}{2g\sqrt{P}}\,,\quad \ext^z=\frac{\sqrt{P}}{\sqrt{H}}\diff z\,,
}
and $x^{\bar\alpha}$ are coordinates on $\R^{1,4}$. We find that the eight Killing spinors preserved by the solution \eqref{circlesol_susy} are subject to the projections
\es{projcirlce}{
\gammasevenL_y\epsilon+\sin\theta\epsilon+\cos\theta\,\gammafiveE_{12}\gammasevenL_z\epsilon=\gammasevenL_y\epsilon+\sin\theta\epsilon+\cos\theta\,\gammafiveE_{34}\gammasevenL_z\epsilon=0\,,
}
where the angle $\theta$ is defined by
\es{thetacircle}{
\sin \theta=\frac{\sqrt{P}}{\pp\,\sqrt{H}}=\frac{2\sqrt{h_+h_-}}{h_++h_-}\,,\quad
\cos\theta=\frac{1}{\pp\,\sqrt{H}}=\frac{h_+-h_-}{h_++h_-}\,,
}
and note that as in the case of spindle solutions we have
\es{thetahalf}{
\sin\tfrac{\theta}{2}=\frac{\sqrt{h_-}}{\sqrt{h_++h_-}}\,,\quad 
\cos\tfrac{\theta}{2}=\frac{\sqrt{h_+}}{\sqrt{h_++h_-}}\,.
}
Solving the differential KSE, one can find an explicit expression for the spinors, which here reads
\es{circlespinors}{
\epsilon=u_+\otimes \chi^{(1)}\otimes \zeta
+u_-\otimes \chi^{(2)}\otimes \zeta^c\,,
}
where $u_{\pm}$ are the two simultaneous eigenvectors of $\gammafiveE_{12}$ and $\gammafiveE_{34}$ introduced in \eqref{upm}, $\chi^{(1,2)}$ are two completely arbitrary 5d spinors (carrying four degrees of freedom each, for a total of eight), and
\es{zetacircle}{
\zeta=
\sqrt{\Omega}\,
\begin{pmatrix}
\cos\tfrac{\theta}{2}\\
-\sin\tfrac{\theta}{2}
\end{pmatrix}\,,
}
while $\zeta^c=B\,(\zeta)^*$ is its 2d charge conjugate introduced previously in this Appendix.

Note that we can decompose the two 4d Dirac spinors $\chi^{(1,2)}$ in positive and negative chirality components $\chi^{(1,2)}_\pm$, with the role of chirality matrix played by $\gammafour_{\rho}$ as in the $\ell=1$ section. We can then write
\es{circlespinornew}{
\epsilon&=\epsilon_+ + \epsilon_-\,,\\
\epsilon_{+}&=u_+\otimes \chi^{(1)}_+\otimes \zeta+u_-\otimes \chi^{(2)}_-\otimes \zeta^c\,,\\
\epsilon_{-}&=u_+\otimes \chi^{(1)}_-\otimes \zeta+u_-\otimes \chi^{(2)}_+\otimes \zeta^c\,,
}
where the two spinors $\epsilon_{\pm}$ (each carrying four degrees of freedom) satisfy
\es{newprojcircle}{
\gammafiveE_{12}\epsilon_\pm\pm \gammasevenL_{\rho y z}\epsilon_\pm=\gammafiveE_{34}\epsilon_\pm\pm \gammasevenL_{\rho y z}\epsilon_\pm=0\,,\quad
(\sin\theta\,\gammasevenL_y\pm\cos\theta\,\gammasevenL_{\rho})\epsilon_\pm+\epsilon_\pm=0\,,
}
which explains the relation with the solutions discussed in Section \ref{sec:7dspindles}.

\section{Deriving the 4d gauged supergravity}\label{app:4dsugraderivation}

In this Appendix we give some details that are useful for the derivation of the action and Killing spinor equations for the abelian truncation of maximal 4d $ISO(7)$ gauged supergravity. The scalars of maximal 4d supergravity parametrize the coset $E_{7(7)}/SU(8)$, where $SU(8)$ is the maximal compact subgroup of $E_{7(7)}$. Other two interesting maximal subgroups of $E_{7(7)}$ are $SL(8,\R)$ and $SL(8,\R)'$, which however are non-compact. For our purposes, there are three useful basis of $E_{7(7)}$, as discussed in \cite{DallAgata:2011aa}: in each of these, the generators of one of the maximal subgroups are diagonal, and the three are related by a change of basis involving $SO(8)$ gamma matrices, where this $SO(8)$ is a common compact subgroup of $SU(8)$, $SL(8,\R)$ and $SL(8,\R)'$. The change of basis essentially permutes triality-equivalent representations of $SO(8)$ and it is described in eq. (2.3) of \cite{DallAgata:2011aa}. $SU(8)$, $SL(8,\R)$ and $SL(8,\R)'$ have dimension $\mathbf{63}=\mathbf{35}+\mathbf{28}$, where $\mathbf{28}$ is the dimension of the adjoint of $SO(8)$, while the dimension of $E_{7(7)}$ is $\mathbf{133}$. One should therefore think of the $\mathbf{70}$ physical scalars of the theory as $\mathbf{35}+\mathbf{35}=\mathbf{70}=\mathbf{133}-\mathbf{63}$, where the two copies of $\mathbf{35}$ correspond to the generators of $SL(8,\R)$ (resp. $SL(8,\R)'$) minus the generators of $SO(8)$. The former are parity even scalars, while the latter are parity odd and will not be relevant here.

All gaugings of 4d maximal supergravity were classified in \cite{deWit:2007kvg} using the embedding tensor, and here we are interested in a purely electric gauging of $ISO(7)\subset SL(8,\R)$. As opposed to the maximal $SO(8)$ gauging, it is important here to specify how $ISO(7)$ is embedded in $E_{7(7)}$. The gauging that we choose here corresponds to the choice $(p,q,r)=(7,0,1)$ in eq. (5.3) of \cite{deWit:2007kvg}, see also \cite{Hull:1988jw} for more comments on $CSO(p,q,r)$ gaugings in 4d. To derive the expressions of the Lagrangian and supersymmetry variations, we find it useful to proceed as follows. We start with a basis where $SL(8,\R)$ (containing the parity-even scalars) is diagonal and identify the four scalars that are invariant under the Cartan generators of $SO(7)\subset ISO(7)$, which are the three gauge fields in our truncation. To do so, we need to introduce some notation. Let us define $SL(8,\R)$ generators $\Lambda$ in the fundamental representations
\es{lambdasl8}{
(\Lambda^A_{\,\,\,B})^C_{\,\,\,D}=\delta^A_D\delta^C_B-\tfrac{1}{8}\delta^A_B\delta^C_D\,,\quad
A,B,\ldots=1,\ldots,8\,,
}
where the $SO(8)$ generators $T$ are identified by the antisymmetric part as
\es{Tso8}{
(T_{AB})_{CD}=\frac{1}{2}(\delta_{AC}\delta_{BD}-\delta_{BC}\delta_{AD})\,,
}
and the position of the indices is unimportant in the $SO(8)$ case. Working in a basis where $SL(8,\R)$ is diagonal has the advantage that the coset representative we are interested in is obtained exponentiating the most general linear combination $\sum_{A=1}^8\lambda_A\Lambda^A_{\,\,\,A}$ of the Cartan generators $\Lambda^A_{\,\,\,A}$ of $SL(8,\R)$ in \eqref{lambdasl8} that is invariant under the $U(1)^3$ generated by $(T_{12},T_{34},T_{56})$ in \eqref{Tso8}. We find that such matrix is given by
\es{}{
H_{\mathbf{8}}=\text{diag}(\lambda_1,\lambda_1,\lambda_2,\lambda_2,\lambda_3,\lambda_3,\lambda_4+4\phi_0,\lambda_4-4\phi_0)\,,
}
where in terms of the scalars appearing in the Lagrangian and KSE we have
\es{}{
\tfrac{1}{2}\lambda_1&=-\phi_1+\phi_2+\phi_3\,,\quad
\tfrac{1}{2}\lambda_2=+\phi_1-\phi_2+\phi_3\,,\\
\tfrac{1}{2}\lambda_3&=+\phi_1+\phi_2-\phi_3\,,\quad
\tfrac{1}{2}\lambda_4=-\phi_1-\phi_2-\phi_3\,,
}
and note that $\sum_{i=1}^4\lambda_i=0$. We can use this to construct the analogue of $H_{\mathbf{8}}$ in the representation $\mathbf{28}$ of $SL(8,\R)$, which is given by
\es{}{
(H_{\mathbf{28}})_{AB}^{\quad CD}=\left[(H_{\mathbf{8}})_{AA}+(H_{\mathbf{8}})_{BB})\right]\delta_{AB}^{CD}\,,\qquad
\delta_{AB}^{CD}\equiv\tfrac{1}{2}(\delta_A^C\delta_B^D-\delta_A^D\delta_B^C)\,.
}
To compare to the expressions in the literature, it is also useful to define
\es{}{
H_{\mathbf{56}}=
\begin{pmatrix}
H_{\mathbf{28}} & 0\\
0 & -H_{\mathbf{28}}^T
\end{pmatrix}\,.
}
In terms of these two objects, we have the coset representative in the $SL(8,\R)$ basis
\es{S_56}{
(L_{\mathbf{56}})_M^{\,\,\,N}=\mathrm{exp}(H_{\mathbf{56}})=
\begin{pmatrix}
(L_{\mathbf{28}})_{AB}^{\quad CD} & 0\\
0 & (L_{\mathbf{28}})^{AB}_{\quad CD}
\end{pmatrix}\,,
}
where $M=1,\ldots,56$ and we are using pairs of fundamental antisymmetric indices $A,B=1,\ldots,8$ for the adjoint of $SL(8,\R)$, which is the notation used in the various papers that we reference \cite{deWit:2007kvg,DallAgata:2011aa,Guarino:2015vca,Guarino:2015qaa,Guarino:2015jca}.

Now that we have defined this object, we explain how it relates to analogous quantities appearing in some of the papers that we reference. First, note that $L_{\mathbf{56}}$ in \eqref{S_56} is in the basis of eq. (2.2) of \cite{DallAgata:2011aa}, which can be written in a mixed $SL(8,\R)-SU(8)$ basis using
\es{mixedbasis}{
(L_{\mathbf{56}})_{M}^{\,\,\,\underline{N}}=(L_{\mathbf{56}})_M^{\,\,\,P}\,(S^{\dagger})_{P}^{\,\,\,\underline{N}}\,\,,
}
where $S_{\underline{M}}^{\,\,\,N}$ is the matrix introduced in eq. (2.3) of \cite{DallAgata:2011aa} implementing the change of basis, and $S^{\dagger}$ is its hermitian conjugate.\footnote{This change of basis was first described in \cite{Hull:1988jw}.} The coset represenative $(L_{\mathbf{56}})_{M}^{\,\,\,\underline{N}}$ in \eqref{mixedbasis} is precisely the one appearing in eq. (3.3) of \cite{deWit:2007kvg}, which we used to derive the supersymmetry variations.\\

Another useful quantity to introduce is the scalar matrix $\mathcal{M}_{AB CD}$ which appears in the uplift formulas \cite{Guarino:2015vca,Guarino:2015qaa,Guarino:2015jca}. This can be conveniently defined in terms of the coset representative $(L_{\mathbf{28}})_{AB}^{\quad CD}$ introduced in \eqref{S_56} as
\es{}{
\mathcal{M}_{ABCD}=(L_{\mathbf{28}}\,L_{\mathbf{28}}^T)_{ABCD}\,.
}
The other fields of the 4d $ISO(7)$ gauged supergravity that we are considering are the three gauge fields called $A_{\mu}^{(I)}$, $I=1,2,3$, in Section \ref{sec:4dsugra}, which in terms of the notation of \cite{Guarino:2015vca,Guarino:2015qaa,Guarino:2015jca} correspond to the three electric gauge fields $\mathcal{A}_{\mu}^{IJ}$ with $(IJ)\in \{(12),(34),(56)\}$, while all other one-forms appearing in {\it e.g.} eq. (3.1) of \cite{Guarino:2015vca} are set to zero. The scalars and gauge fields described above are enough to determine the dilaton and metric of type IIA, while the other field that is non-trivial in the uplifted solution is the RR four-form flux $F_{(4)}$,  or its potential $A_{(3)}$,which is given in eq. (3.12) of \cite{Guarino:2015vca}. The only non-trivial fields in that expression, for our specific case, are the three-form $\mathcal{C}^{IJ}$ and the two-form $\mathcal{B}_J^{\,\,\,I}$, which can be determined using eqs. (3.29-3.33) of \cite{Guarino:2015vca} (as well as the definitions in Section 2 of \cite{Guarino:2015qaa}.

\newpage

\providecommand{\href}[2]{#2}\begingroup\raggedright\endgroup

\end{document}